\documentclass[twocolumn,showpacs,preprintnumbers,amsmath,amssymb,superscriptaddress]{revtex4}

\usepackage{graphicx}
\usepackage{epstopdf}
\usepackage{dcolumn}
\usepackage{bm}

\begin{document}


\title{ 
Recent advances of NMR and NQR studies of 
YBa$_2$Cu$_3$O$_{7-\delta}$ and YBa$_2$Cu$_4$O$_8$
}
\author{
Y. Itoh
}

\address{
Department of Chemistry, Graduate School of Science, 
Kyoto University, 606-8502, Japan
}


\begin{abstract}
Nuclear magnetic resonance (NMR) and nuclear quadrupole resonance (NQR) techniques 
have supplied us with unique information on local incoherent phenomenon
as well as low frequency magnetic response of electronic systems. 
Some of recent advances in the studies of high-$T_\mathrm{c}$ cuprate superconductors are based on 
site-selective measurements of the local electronic state and the local electron spin dynamics. 
High-$T_\mathrm{c}$ superconductors YBa$_2$Cu$_3$O$_{7-\delta}$ (the optimal $T_\mathrm{c}$ = 93 K) 
and YBa$_2$Cu$_4$O$_8$ ($T_\mathrm{c}$ = 82 K) with double-layer CuO$_2$ planes have been studied by NMR and NQR techniques. 
From the intensive studies at the early stage, 
two-dimensional antiferromagnetic spin fluctuations and pseudo spin-gap in the magnetic excitation spectrum 
turned out to be key ingredients to characterize the novel electronic states, associated with the occurrence of superconductivity. 
Wipeout effect is a new key to characterize superconductor-insulator boundary in the lightly doped regime and impurity substitution effects.  
What happens near in-plane impurities and inside vortex cores in such novel electronic states? 
Nonmagnetic impurity Zn decreases $T_\mathrm{c}$ of the $d$-wave superconductivity because of the breakdown of Anderson theorem.
From the careful NMR and NQR studies, it was concluded that the pseudo spin-gap in the normal state is robust for the substitution of nonmagnetic impurity Zn and magnetic impurity Ni. 
In the vicinity of a superconductor-insulator boundary induced by impurity, however, the pseudo spin-gap is suppressed.
Collapse of low energy pseudo spin-gap due to the heavily substituted impurity Zn was also evidenced by neutron scattering measurements. 
For the impurity-substituted samples and in the vortex state of pure samples,
the characteristic NMR and NQR spectra were observed.  
The local enhancement of magnetic correlation near the impurity Zn and inside the vortex core 
was observed from the frequency distribution of nuclear spin-lattice relation time $T_1$. 
Scanning tunneling spectroscopy detects local density of states of electrons near Zn in the superconducting state. 
NMR and NQR experiments detect the local magnetism near Zn  and inside the vortex cores. 
\end{abstract}

\maketitle

\section{INTRODUCTION}
\label{sec:intro}
Nuclear magnetic resonance (NMR) and nuclear quadrupole resonance (NQR) 
are powerful techniques to characterize local incoherent phenomenon in materials~\cite{Slichter,Slichter2007}, 
whereas neutron and X-ray scattering techniques are sensitive to coherent phenomenon~\cite{Marshall}. 
Microscopic studies using the NMR and NQR techniques have provided us with rich information  
on crystalline imperfection, dilute exchange spin networks, Friedel oscillation, Ruderman-Kittel-Kasuya-Yosida (RKKY)
oscillation, and Kondo screening effect in materials~\cite{Narath,Jac,Alloul,Gruner}.  
Some of recent advances in the NMR and NQR studies of high-$T_\mathrm{c}$ cuprate superconductors are based on 
site-selective measurements of the local electronic state and the local electron spin dynamics.  
\begin{figure}[h]
\begin{center}
\includegraphics[width=0.9\linewidth]{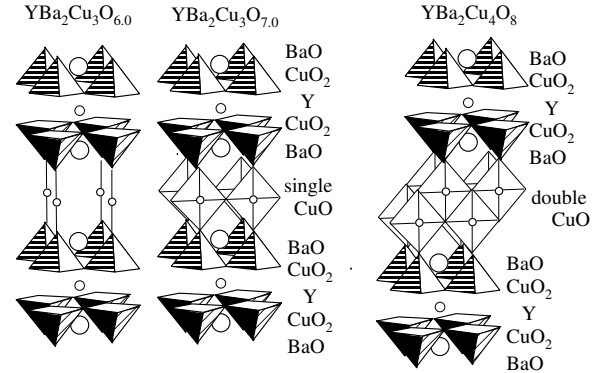}
\end{center}
\caption{
Crystal structures of pure undoped YBa$_2$Cu$_3$O$_{6.0}$, 
optimally doped superconductor YBa$_2$Cu$_3$O$_{7}$, and naturally underdoped superconductor YBa$_2$Cu$_{4}$O$_{8}$. 
}
\label{CS}
\end{figure} 
\begin{figure}[h]
\begin{center}
\includegraphics[width=0.8\linewidth]{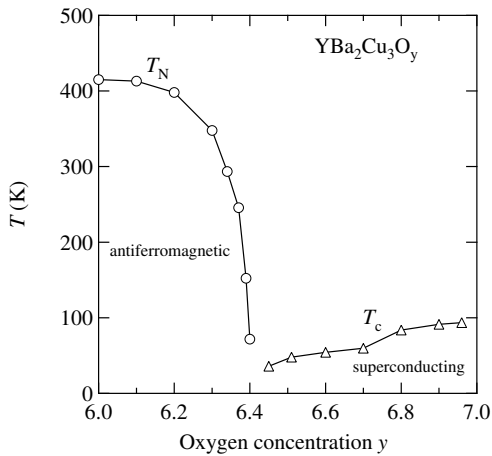}
\end{center}
\caption{
Phase diagram of Y123y reproduced from Rossat-Mignod {\it et al}.~\cite{Rossat}. 
$T_\mathrm{N}$ and $T_\mathrm{c}$ are plotted against oxygen concentration $y$.
}
\label{PD}
\end{figure} 

YBa$_2$Cu$_3$O$_y$ (6.4$\leq y \leq$7) (Y123$y$) system is a widely known high-$T_\mathrm{c}$ superconductor. 
Y123$y$ has two inequivalent crystallographic Cu sites;
 bilayer plane-site Cu(2) and single chain-site Cu(1).  
Figure~\ref{CS} shows the crystal structures of the Y123 family. 
The maximum superconducting transition temperature $T_\mathrm{c}$ is 93 K
at the optimal oxygen concentration of $y$ = 6.92$\sim$6.96. 
For oxygen deficiency of  6.4$\leq y \leq$6.6, 
the $T_\mathrm{c}\sim$ 60 K phase is known to show a typical underdoped electronic state
with pseudo gap and pseudo spin-gap.  
Figure~\ref{PD} is a typical phase diagram of the Y123~\cite{Rossat}. 

Non-stoichiometry in compounds gives rise to inhomogeneous broadening in NMR and NQR spectra,
which often prevents us from extracting intrinsic electronic properties. 
YBa$_2$Cu$_4$O$_8$ (Y124) with bilayer CuO$_{2}$ planes and double CuO chains
has $T_\mathrm{c}$ = 82 K.
It is a stoichiometric and naturally underdoped compound. 
The oxygen concentration is hard to be removed.   
Because of chemical stability and of lack of oxygen disorder effects, 
since the first observations of Cu NQR~\cite{Mali,Zimmermann}, 
NMR and NQR studies have been intensively carried out for Y124 as well as the Y123 family
to reveal intrinsic electronic properties~\cite{PS,Berthier,Rigamonti}. 

The aim of this article is to present some of the recent advances of NMR and NQR studies of high-$T_\mathrm{c}$ superconductors
Y123 and Y124. The article is organized to show  
(II) the present understanding of magnetic phase diagram with respect to doped hole concentration, 
(III) impurity-substitution effects, 
(IV) impurity-induced NMR and NQR relaxation, 
and (V) vortex core state in the superconducting mixed state. 
Recent scanning tunneling microscopy (STM) and spectroscopy (STS) studies of high-$T_\mathrm{c}$ superconductors have supplied us with vital information of local density of states of electron\cite{Balatsky,Fischer}.   
Local microscopic magnetism near impurity and vortex core, which we are concerned about, is the matter
that NMR and NQR can clarify actually.    

\section{MAGNETIC PHASE DIAGRAM}
\subsection{TWO DIMENSIONAL RENORMALIZED CLASSICAL REGIME}

The parent undoped compound YBa$_2$Cu$_3$O$_6$ is a Mott insulator and bilayer-coupled  
Heisenberg antiferromagnet with a N{\' e}el temperature $T_\mathrm{N}$ = 415 K. 
The plane-site Cu(2)$^{2+}$ carries an unpaired moment $S$ = 1/2. 
The chain-site Cu(1)$^{1+}$ has a closed shell. 
The paramagnetic state is in two dimensional renormalized classical regime,
where the antiferromagnetic correlation length $\xi$ diverges exponentially 
with cooling toward $T$ = 0 K~\cite{AA,CNN,MT} as
\begin{equation}
\xi/a \approx 0.5\mathrm{exp}(2\pi \rho_s/T),
\label{RCxi}
\end{equation}
where $\rho_s$ is a spin stiffness constant proportional to an in-plane exchange coupling constant $J$. 
The value of $J$ is estimated to be about 150 meV~\cite{RM}.
 Because of a spin $S$ = 1/2 system, there is no single-ion anisotropy
 due to electrostatic crystalline potential. 
 The spin wave dispersion in the N{\' e}el  state is nearly gapless~\cite{RM},
which leads to high Cu nuclear spin-lattice relaxation rate at low temperatures below $T_\mathrm{N}$~\cite{Tsuda}. 
Thus, the $T_\mathrm{N}$ higher than room temperature can be considered to result from 
the strong in-plane $J$ with XY anisotropy~\cite{Hanzawa} and three dimensional interactions of $J_{\perp}$
as $k_\mathrm{B}T_\mathrm{N}\approx J_{\perp}(\xi/a)^2$ ~\cite{CNN}.

\begin{figure}[h]
\begin{center}
\includegraphics[width=0.8\linewidth]{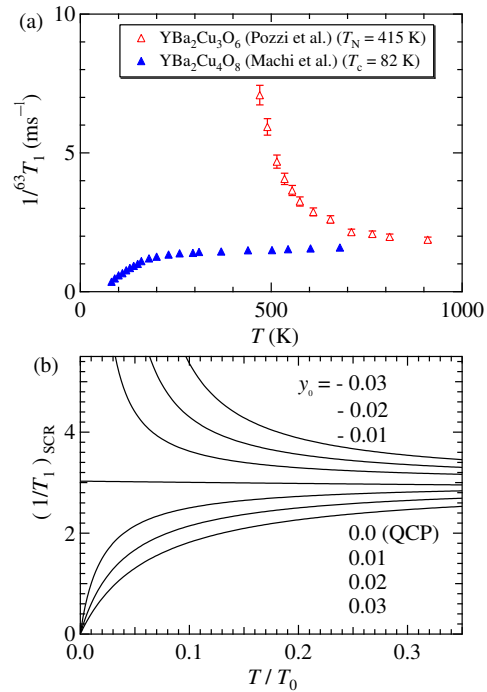}
\end{center}
\caption{
(a) High temperature $^{63}$Cu nuclear spin-lattice relaxation rate 1/$T_{1}$ 
of parent insulator YBa$_2$Cu$_3$O$_6$ reproduced from Pozzi {\it et al}.~\cite{Pozzi} and 
of naturally undersoped YBa$_2$Cu$_4$O$_8$ reproduced from Machi {\it et al}.~\cite{MachiImai}.
(b) The nuclear spin-lattice relaxation rate 1/$T_1$ calculated by SCR (self-consistent renormalization)
theory from weakly antiferromagnetic to nearly antiferromagnet in two dimension reproduced from~\cite{MTU}.  
$T_\mathrm{0}$ is a spin fluctuation energy.  $T/T_\mathrm{0}$ is a reduced temperature. 
}
\label{RC}
\end{figure} 

The renormalized classical behavior in Cu nuclear spin-lattice relaxation was first observed by Imai {\it et al.}~\cite{Imai} for La$_2$CuO$_4$. 
The measurement of high temperature spin dynamics is a highlight in the successful NMR spin-echo experiments.  
Subsequently, such a renormalized classical behavior in NMR relaxation was observed by Pozzi {\it et al}. 
for Y1236~\cite{Pozzi}, Ca$_{0.85}$Sr$_{0.15}$CuO$_{2}$ ($T_\mathrm{N}$ = 539 K)~\cite{Pozzi2},
by Fujiyama {\it et al.} for PrBa$_2$Cu$_4$O$_8$ ($T_\mathrm{N}$ = 230 K)~\cite{Fujiyama} and 
by Thurber {\it et al.} for three-leg ladder Sr$_2$Cu$_3$O$_5$~\cite{ImaiL}.

Figure~\ref{RC} (a) shows the high temperature $^{63}$Cu nuclear spin-lattice relaxation rate 1/$T_{1}$ 
of YBa$_2$Cu$_3$O$_6$ reproduced from Pozzi {\it et al}.~\cite{Pozzi} and 
of naturally underdoped YBa$_2$Cu$_4$O$_8$ reproduced from Machi {\it et al}.~\cite{Machi}.

\subsection{SUPERCONDUCTOR-INSULATOR BOUNDARY}

\begin{figure}[h]
\begin{center}
\includegraphics[width=0.9\linewidth]{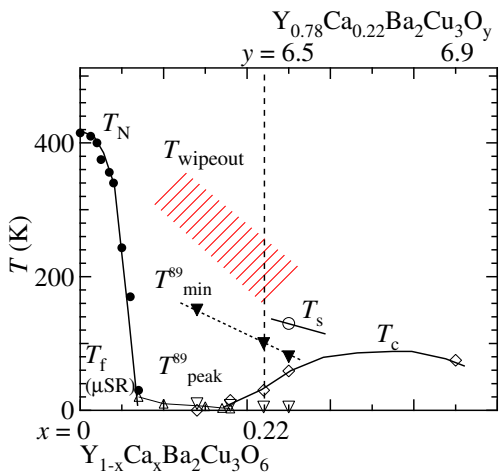}
\end{center}
\caption{
Magnetic phase diagrams of Y$_{1-x}$Ca$_x$Ba$_2$Cu$_3$O$_y$
reproduced from Singer and Imai~\cite{Singer}. 
}
\label{YCaMP}
\end{figure} 

As the hole concentration increases, $T_\mathrm{N}$ is reduced
and finally the superconductivity emerges. 
According to a field theory applied to a non-linear sigma model~\cite{CNN,ChubSach,Sandvik} and to a modified spin wave theory~\cite{Yosida} in the insulating regime, 
a spin frustration effect destroys the coherency of an antiferromagnetic network
and finally a paramagnetic state or disordered gapped state should emerge
through a two dimensional quantum critical point (QCP). 
According to the SCR (self-consistent renormalized) theory for a two dimensional spin fluctuation model 
in the itinerant regime, 
the static staggered magnetisation in a weakly antiferromagnetic state is reduced and 
finally a nearly antiferromagnetic state emerges
through an itinerant two dimensional QCP~\cite{MTU}.  

Figure~\ref{RC} (b) shows the nuclear spin-lattice relaxation rate calculated by SCR (self-consistent renormalization)
theory from weakly antiferromagnetic to nearly antiferromagnet in two dimension reproduced from~\cite{MTU}.    
Although the development of the spin dynamics with hole doping from Y1236 to Y1248 in Fig.~\ref{RC} (a) 
is similar to that via the QCP in Fig.~\ref{RC} (b),  
the actual cuprates change from insulator to superconductors through a Mott transition. 
What happens at the boundary from insulating side to superconducting side?

Wipeout effect on NMR is a key to characterize the physics at the boundary of the Y123 family
as well as La$_{2-x}$Sr$_x$CuO$_4$~\cite{Imai1,Imai2,JB}. 
Sudden disappearance of Cu NMR and NQR signals is observed 
near the superconductor-to-insulator boundary of Y$_{1-x}$Ca$_x$Ba$_2$Cu$_3$O$_y$.
Figure~\ref{YCaMP} shows a magnetic phase diagram of Y$_{1-x}$Ca$_x$Ba$_2$Cu$_3$O$_y$~\cite{Singer}.
In general, the wipeout is recognized as an ill-defined physical quantity. 
Nevertheless, it well characterizes what happens near the phase boundary. 
An abrupt wipeout effect is different from conventional slowing down phenomenon,
where the nuclear spin relaxation should become fast.
The wipeout effect without any slowing down effect on the observable NMR/NQR signals
leads us to conclude that
the electron spin fluctuation spectrum should consist of two components in frequency space. 
One is a main spectrum, which may be called a {\it fast} mode.
The other is the induced low frequency spectrum, similar to a central mode in structural phase transition,
which may be called a {\it slow} mode.   
In Fig.~\ref{YCaMP}, 
the existence of the slow mode is evidenced by ligand $^{89}$ Y NMR relaxation measurements~\cite{Singer}. 
Because of a weak $^{89}$Y nuclear-electron coupling constant~\cite{AOM}, 
the fast nuclear relaxation component being unobservable by Cu nuclear spin-echo 
is observed by  $^{89}$ Y NMR. 
 
\begin{figure}[h]
\begin{center}
\includegraphics[width=0.8\linewidth]{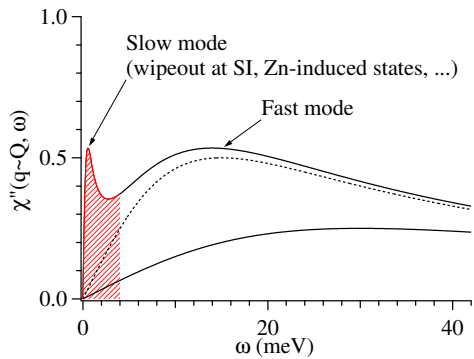}
\end{center}
\caption{
Two components of Cu electron spin fluctuation spectrum. 
The slow mode is inferred from wipeout effect on NMR and NQR.
Such a mode near superconductor-insulator boundary 
might be associated with Zhang-Rice bound singlets.  
}
\label{WP}
\end{figure} 
The spin fluctuation spectrum with two energy scales reminds us of 
a phenomenological itinerant-localized dual model~\cite{Miyake,Narikiyo}. 
The actual slow mode is inhomogeneous in space and in frequency.  
In most cases, Mott transition is first order transition. 
Thus, such a slow mode may be associated with the character of first order transition.  

Holes are doped into the plane-site oxygens~\cite{IFT}. 
In the superconducting regime, the doped holes do not have independent spin degree of freedom, 
which is indicated by a single scaling relation of $^{17}$O(2, 3) with $^{63}$Cu(2) Knight shifts~\cite{Takigawa}.  
The leading spin component of conduction band is single~\cite{AOM,Takigawa,Imai3}. 
The doped holes do not carry isolated free moments  but are considered to form Zhang-Rice bound singlets of 3$d^{9}\b{L}$~\cite{ZR}.  
The Zhang-Rice singlets are localized in the semi-conducting regime at low temperatures. 
The phenomenological slow mode may be associated with the localized Zhang-Rice singlets,
which cause the wipeout effect on NMR/NQR.  

Can we observe NMR/NQR signals associated with the Zhange-Rice singlets
in any cases?
One may find a few examples of high frequency Cu NQR signals in some doped cuprates. 
For the hole-doped ladder and linear chain oxides of Sr$_{14}$Cu$_{24}$O$_{41}$~\cite{TakigawaL} 
and Ca$_{0.85}$CuO$_{2}$~\cite{Yokoyama}, 
Cu NMR/NQR signals near the Zhang-Rice singlets are observed in the chain sites. 
In spite of the singlets, the nuclear spin-lattice relaxation time is relatively short. 
For Li-substituted La$_2$CuO$_4$~\cite{Rykov} and NaCuO$_{2}$~\cite{Luders},  
non-mganetic Cu$^{3+}$ NQR signals are observed to be in low spin electronic states.  
The observed magnetism of Cu$^{2+\delta}$ of 3$d^{9}\b{L}$ is different from Cu$^{3+}$ so far.
One should proceed further site-selected NMR studies to observe 
the local electronic state of Zhang-Rice singlets in doped oxides. 

\subsection{NORMAL-STATE PSEUDO SPIN-GAP}
Normal-state pseudo gap and pseudo spin-gap are characteristic anomalies of the underdoped 
superconductors.  
The existence of normal state gap is inferred from the above-$T_\mathrm{c}$ decrease of nuclear spin-lattice relaxation rate divided by temperature 1/$T_{1}T$~\cite{ImaiPSG,AOM,Takigawa,YasuokaPSG,Warren,Horvatic}.  
1/$T_{1}T$ is the square of hyperfine coupling constant $A^2$ times the wave vector averaged low frequency dynamical spin susceptibility $\chi "({\bf q}, \omega)$ (wave vector $\bf q$ and frequency $\omega$)~\cite{Moriya0}. 
Inelastic neutron scattering confirms the low frequency suppression of $\chi "({\bf q}, \omega)$ 
for underdoped Y123~\cite{RM}. 
The effect of the pseudo spin-gap is also recognized in transport properties~\cite{Uchida}.   
\begin{figure}[h]
\begin{center}
\includegraphics[width=0.8\linewidth]{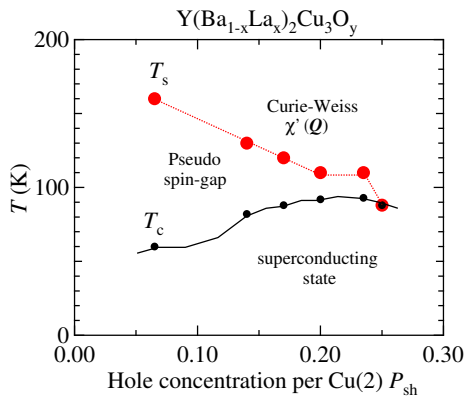}
\end{center}
\caption{
Pseudo spin-gap of Y(Ba$_{1-x}$La$_x$)$_2$Cu$_3$O$_y$
reproduced from Matsumura {\it et al}.~\cite{Matsumura} and Auler {\it et al}.~\cite{Auler}.  
The hole concentration $P_\mathrm{sh}$ per CuO$_2$ sheet is defined by Cu$^{2+P_\mathrm{sh}}$. 
}
\label{SG}
\end{figure} 
\begin{figure}[h]
\begin{center}
\includegraphics[width=0.8\linewidth]{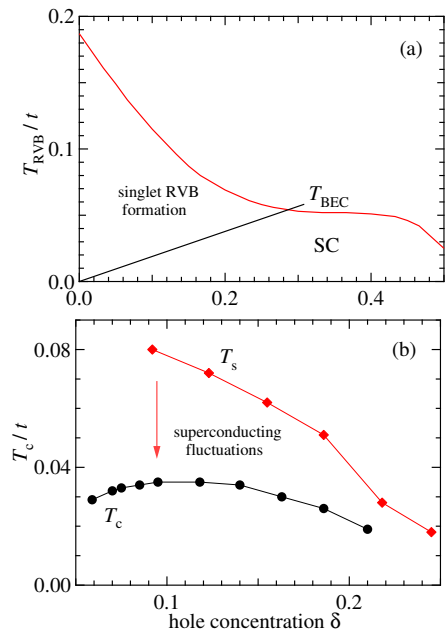}
\end{center}
\caption{
Hole doping dependence of pseudo spin-gap temperature $T_\mathrm{s}$ 
as a siglet-RVB formation temperature $T_\mathrm{RVB}$ in two dimensional $t$-$J$
model factorized by spinon and holon operators reproduced from Tanamoto {\it et al}.~\cite{TKF} (a), and  as an onset temperature (a mean-field $T_\mathrm{c}^{\mathrm{MF}}$) of $d$-wave superconducting fluctuations reproduced from Yanase and Yamda~\cite{Yanase} (b).
}
\label{theorySG}
\end{figure} 

Figure~\ref{SG} shows the typical hole doping dependence of pseudo spin-gap temperature $T_\mathrm{s}$
for Y(Ba$_{1-x}$La$_x$)$_{2}$Cu$_{3}$O$_{y}$,
which is defined by the maximum temperature of Cu 1/$^{63}T_{1}T$~\cite{Matsumura,Auler}. 
 The effect of the opening of pseudo spin-gap is not a phase transition but a crossover phenomena. 
The $T_\mathrm{s}$ decreases as the hole concentration per Cu increases and then
coincides with $T_\mathrm{c}$ at slightly overdoped concentration.  
Although the existence of overdoped regime of Y123y is known for single crystals,
it was unclear for polycrystalline samples. 
Yamamoto {\it et al.} found the method of a solid state reaction and heat treatment 
to synthesize the overdoped polycrystalline Y1237 with $T_\mathrm{c}$ = 88 K~\cite{Yamamoto}. 

One should note that  the above-$T_\mathrm{s}$ behavior of 1/$^{63}T_{1}T$
is of Curie-Weiss type but not Korringa law~\cite{Singer,WilliamsOV}. 
Although the non-Korringa behavior in the overdoped regime is not widely recognized, 
the Curie-Weiss type 1/$^{63}T_{1}T$ is also confirmed 
for overdoped La$_{1.80}$S$_{0.20}$CuO$_{4}$~\cite{Itoh020}, 
La$_{1.77}$S$_{0.23}$CuO$_{4}$~\cite{Itoh023}, 
and HgBa$_2$CuO$_{4+\delta}$~\cite{ItohOV}. 
The high temperature electronic state in the overdoped regime is not 
conventional Landau-Fermi liquid. 
There is no phase boundary in the high temperature electronic state 
at the optimally doped-to-overdoped regimes. 
 
Since the pseudo spin-gap effect was not obvious in  single layer La$_{2-x}$Sr$_2$CuO$_4$, 
it had been suspected that such a gap is not essential for high $T_\mathrm{c}$.  
Does the pseudo spin-gap result from a spin singlet paring in the adjacent CuO$_2$ layers as a property peculiar to double layer compounds~\cite{MM} or from an inherent property in a single CuO$_2$ plane? 
It had been a long term problem. 
Distinct conclusion was obtained from single layer HgBa$_2$CuO$_{4+\delta}$~\cite{ItohHg1201,BobroffHg1201}.
The pseudo spin-gap is an inherent nature of the high-$T_\mathrm{c}$ superconductors,
irrespectively of the number of CuO$_2$ planes per unit cell. 
The doping dependence of $T_\mathrm{s}$ of the single layer HgBa$_2$CuO$_{4+\delta}$ is different from 
that of the double layer HgBa$_2$CaCu$_2$O$_{6+\delta}$~\cite{ItohHg1212}. 
The $T_\mathrm{s}$ as a function of the hole concentration depends on the number of CuO$_2$ planes per unit cell
and on the shape of underlying electronic Fermi surface. 

Two theoretical explanations have been proposed.
In the two dimensional $t$-$J$ model with spinon-holon decomposition technique,
the pseudo spin-gap temperature $T_\mathrm{s}$ is regarded as the onset temperature $T_\mathrm{RVB}$ of singlet RVB (resonating valence bond) formation of spinons~\cite{Kotliar,Suzumura,TKF}. 
The real $T_\mathrm{c}$ is given by Bose-Einstein condensation temperature $T_\mathrm{BEC}$,
leading to the underdoped regime. 
In the existing calculations, $T_\mathrm{RVB}$ is given by a second order phase transition temperature. 
Figure~\ref{theorySG} (a) shows a numerical calculation of $T_\mathrm{RVB}$ as a function of
doped hole concentration~\cite{TKF}. 
The doping dependence of $T_\mathrm{RVB}$ depends on the contour of a basal Fermi surface. 
For each high-$T_\mathrm{c}$ family with different Fermi surface, $T_\mathrm{RVB}$ shows a different doping dependence. 
 
In the two dimensional superconducting fluctuation theory,   
the pseudo spin-gap temperature $T_\mathrm{s}$ is regarded as the onset of enhancement
of $d$-wave superconducting fluctuations and a mean-field $T_\mathrm{c}$. 
The actual $T_\mathrm{c}$ is reduced by the strong superconducting fluctuations
so that the underdoped regime appears.
Thus,  the mean-field $T_\mathrm{c}$ is a crossover temperature. 
Figure~\ref{theorySG} (b) shows the mean-field $T_\mathrm{c}$ and the suppressed $T_\mathrm{c}$~\cite{Yanase} 

\section{IMPURITY-SUBSTITUTION EFFECT}
\subsection{PAIR BREAKING AND SUPERUNITARY}
Magnetic and nonmagnetic impurity substitution effects have been studied 
for the high-$T_\mathrm{c}$ superconductors,
where the impurity is substituted for the plane-site Cu(2). 
The impurity Zn (spinless $S$ = 0) induces Curie magnetism, a local staggered spin susceptibility
in the normal state,
an in-gap antiferromagetic dynamical spin susceptibility, and 
a virtual bound electronic state around Zn in the
superconducting state.  

In strong coupling superconductors, 
the actual super conducting transition temperature $T_\mathrm{c}$ 
is determined by competition between pairing effect and depairing effect~\cite{OhashiShiba}. 
In the phonon-mediated superconductors, 
the depairing effect due to thermal phonon  competes the paring effect due to virtual phonon.
For $s$-wave superconductors, 
non-magneitc impurities do not suppress superconductivity because of Anderson's theorem,
unless they induce Anderson localization effect. 
For $d$-wave superconductors, 
non-magneitc impurities suppress superconductivity.
This is a natural consequence of breakdown of Anderson's theorem~\cite{Millis,Balian}.  
The superconducting state of the high-$T_\mathrm{c}$ superconductors 
is characterized by $d_{x^2-y^2}$-wave pairing~\cite{Scalapino}.   

The criterion of scattering strength of impurities is given by the value of 
$\gamma_{n, m}$,  
\begin{eqnarray}
\left\{
\begin{array}{l}
\gamma_n=\pi N_\mathrm{F}u,\\
\gamma_m=\pi N_\mathrm{F}JS,
\end{array}
\right.
\label{uJS}
\end{eqnarray}
where $N_\mathrm{F}$ is the density of states of electrons at the Fermi level,
a nonmagnetic potential depth $u$, and an exchange coupling constant $J$ of a magnetic impurity $S$ with conduction electrons. 
A weak scatterer can be treated by lowest order Born approximation as $\left| \gamma_{n, m}\right|<$ 1.
A strong scatter can be  treated by unitarity limit as $\left| \gamma_{n, m}\right|>$ 1.    

In an $s$-wave superconductor, weak magnetic scatterers suppress superconductivity:
\begin{equation}
\Delta T_\mathrm{c}=0.5\pi^2c_\mathrm{imp}N_\mathrm{F}J^2S(S+1)/k_\mathrm{B},
\label{sWTc}
\end{equation}
where $c_\mathrm{imp}$ is in-plane impurity concentration. 
This is the Abrikosov-Gorkov's formula.
In a $d$-wave superconductor, weak magnetic scatterers suppress superconductivity by
\begin{equation}
\Delta T_\mathrm{c}=0.25\pi^2c_\mathrm{imp}N_\mathrm{F}J^2S(S+1)/k_\mathrm{B}.
\label{dWTc}
\end{equation}
In the lowest order Born approximation, the $d$-wave superconductor
is twice robust than the $s$-wave superconductor for the destructive magnetic impurities~\cite{Maekawa,Openov}. 

For the magnetic impurity effect in $s$-wave superconductors,
one can neglect nonmagnetic potential scattering 
with each magnetic impurity, because of Anderson's theorem. 
However, the potential scattering also affects $d$-wave superconductivity 
as well as the magnetic impurity scattering. 
Thus one has to take into account both effects~\cite{Maekawa,Salkola,Kilian}. 

Taking into account the impurity scatterings of classical spins ($S_z=\pm S$) 
within a $t$-matrix approximation, Ohashi derived the theoretical expression of $T_\mathrm{c}$ ($c_\mathrm{imp}\ll$1) in a $d_{x^2-y^2}$-wave
superconductor~\cite{ItohNi},
\begin{equation}
\Delta T_\mathrm{c}={c_{\rm imp} \over 4k_\mathrm{B}N_\mathrm{F}}
\bigl[
1-{1 \over 2}
(
{1 \over 1+(\gamma_n-\gamma_m)^2}+
{1 \over 1+(\gamma_n+\gamma_m)^2}
)
\bigr].
\label{OhashiTc}
\end{equation}
For $\gamma_m=0$, 
eq.(\ref{OhashiTc}) gives $\Delta T_\mathrm{c}$ in the case of nonmagnetic impurity Zn
~\cite{Borkowski,Fehrenbacher,Arberg,Kitaoka}, 
whereas $\gamma_n=0$ gives $\Delta T_\mathrm{c}$ in the case of magnetic impurity without nonmagnetic potential
scattering. 
In the limit of $\gamma_m\rightarrow 0$ and $\gamma_n\rightarrow 0$, 
eq.(\ref{OhashiTc}) leads to the expression of $\Delta T_\mathrm{c}$ in the lowest
order Born approximation~\cite{Maekawa}. 
In addition, $\gamma_m\to\infty$ or $\gamma_n\to \infty$ corresponds to the unitarity limit~\cite{Hotta}:
 \begin{equation}
\Delta T_\mathrm{c}={c_\mathrm{imp} \over 4k_\mathrm{B}N_\mathrm{F}}. 
\label{ZnTc}
\end{equation}  
If cancellation of $\gamma_m-\gamma_n=0$ in the unitarity limit occurs for Ni, 
one obtains 
\begin{equation}
\Delta T_\mathrm{c}={c_\mathrm{imp} \over 8k_\mathrm{B}N_\mathrm{F}}, 
\label{NiTc}
\end{equation}
which is two times smaller than that due to pure nonmagnetic scattering.

\begin{figure}[h]
\begin{center}
\includegraphics[width=0.8\linewidth]{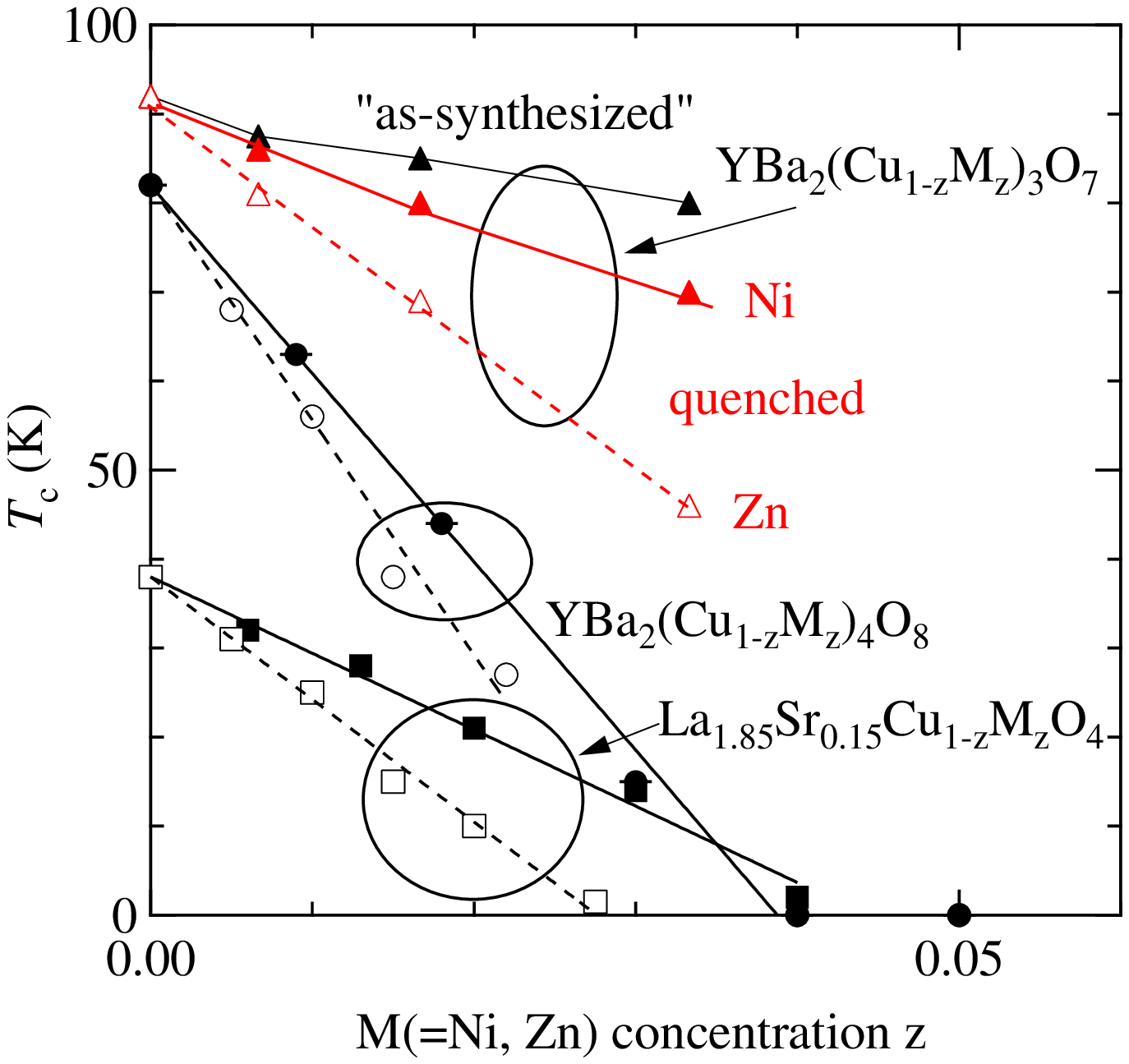}
\end{center}
\caption{
$T_\mathrm{c}$ versus impurity concentration $z$ for various high-$T_\mathrm{c}$ superconductors
with M=Ni or Zn.
The solid (open) symbols are the Ni(Zn) doping dependence of $T_\mathrm{c}$.
The data are adopted from~\cite{AdachiNi,AdachiZn} for YBa$_2$(Cu$_{1-z}$M$_z$)$_3$O$_7$ (upward triangles), 
~\cite{Itoh,Miyatake,Watanabe} for YBa$_2$(Cu$_{1-z}$M$_z$)$_4$O$_8$ (circles), 
and~\cite{Xiao} for La$_{1.85}$Sr$_{0.15}$Cu$_{1-z}$M$_z$O$_4$ (downward triangles). 
The decrease of $T_\mathrm{c}$ by Ni doping for Y123 is most slow among these compounds.
Ni doping dependence of $T_\mathrm{c}$ for ``as-synthesized" or ``quenched" Y123, adopted from~\cite{AdachiNi}. 
The solid curves are guide for the eyes. 
The decrease of $T_\mathrm{c}$ by Ni doping is larger in ``quenched" Y123 than in ``as-synthesized" one. 
}
\label{impTc}
\end{figure}
In Fig.~\ref{impTc}, impurity substitution effects on $T_\mathrm{c}$ are plotted against the concentration $z$ 
for various high-$T_\mathrm{c}$ superconductors with impurity M=Ni or Zn.
The solid (open) symbols are the Ni(Zn) doping dependence of $T_\mathrm{c}$.
The data are adopted from~\cite{AdachiNi,AdachiZn}  
for optimally doped YBa$_2$(Cu$_{1-z}$M$_z$)$_3$O$_7$ (upward triangles), 
~\cite{Itoh,Miyatake,Watanabe} for naturally underdoped YBa$_2$(Cu$_{1-z}$M$_z$)$_4$O$_8$ (circles), 
and~\cite{Xiao} for La$_{1.85}$Sr$_{0.15}$Cu$_{1-z}$M$_z$O$_4$ (downward triangles). 
The degree of suppression of $T_\mathrm{c}$ by Zn and Ni impurities 
depends on the hole doping level in Y123y~\cite{Liang}. 
For less concentration of oxygen, 
$\Delta T_\mathrm{c}/\Delta c_\mathrm{imp}$ becomes larger. 

The decrease of $T_\mathrm{c}$ by Ni substitution for the optimally doped Y1237 is most slow among these compounds.
Using a reduced oxygen partial pressure technique, Adachi {\it et al}. succeeded in controlling 
$T_\mathrm{c}$ of YBa$_2$Cu$_{3-x}$Ni$_x$O$_{7-\delta}$ with optimal oxygen concentration~\cite{AdachiNi}. 
They synthesized two series of YBa$_2$Cu$_{3-x}$Ni$_x$O$_{7-\delta}$ 
samples with the different $T_\mathrm{c}$'s per Ni concentration. 
We call the samples with $T_\mathrm{c}$ denoted by black upward triangles 
``{\it as-synthesized}" ones, because they were synthesized in flowing oxygen gas without quenching treatment, 
and the other samples  with lower $T_\mathrm{c}$ denoted by red upward triangles 
``{\it quenched}" ones,
because they were the as-synthesized samples once again fired and quenched in a reduced 
oxygen atmosphere at 800 $^\circ$C. 
In Fig.~\ref{impTc}, 
Ni substitution dependence of $T_\mathrm{c}$ is plotted for ``as-synthesized" or ``quenched" Y123~\cite{AdachiNi}.
The decrease of $T_\mathrm{c}$ by Ni is larger in ``quenched" Y123 than in ``as-synthesized" one. 
Ni prefers the higher coordination of oxygen atoms. 
The plane-site Cu(2) is located in the pyramid with five oxygen ions, 
whereas the chain-site Cu(1) is coordinated with two, three, or four nearest neighbor oxygen ions. 
The two series of samples with different $T_\mathrm{c}$ 
suggest that the distribution of Ni-substituted sites over Cu(1) and Cu(2) sites 
is changed through synthesis under the reduced oxygen atmosphere.  

From the Cu(1) and Cu(2) NQR measurements, 
it turned out that the heat treatment
in reduced oxygen atmosphere results in a redistribution of the Ni ions 
in YBa$_2$Cu$_{3-x}$Ni$_x$O$_{7-\delta}$~\cite{ItohNi}.
The plane-site $^{63}$Cu(2) nuclear spin-lattice relaxation for the quenched sample was faster
than that for the as-synthesized sample, in contrast to the $^{63}$Cu(1) relaxation that was faster for the as-synthesized sample.
This indicates that the density of plane-site Ni(2) is higher in the quenched samples, 
contrary to the chain-site Ni(1) density which is lower in the quenched samples. 
The Ni substitution for the chain Cu(1) site is a reason 
why the suppression of $T_\mathrm{c}$ by Ni impurity is so slow in the optimally doped Y1237.  
In passing, Co impurity is also substituted both for Cu(1) and Cu(2) sites in the 1237,
which was verified by Cu(1) and Cu(2) NQR by Kohori {\it et al}.~\cite{Kohori} prior to the Ni study~\cite{ItohNi}.

Monthoux and Pines had an insight into the impurity substitution effect on the host spin fluctuation
spectrum, which yields the pairing interaction of high $T_\mathrm{c}$,  
as well as the life time of superconducting quasi-particles of Y123~\cite{Monthoux}. 
They pointed out that the observed suppression of $T_\mathrm{c}$ per Zn 
is too strong to be ascribed only to a simple potential scattering effect on quasi-particle lifetime:
\begin{equation}
(\Delta T_\mathrm{c})_\mathrm{unitary} \approx 5 (\mathrm{K}/\%)\times (100c_\mathrm{imp}). 
\label{SuperUnitary}
\end{equation}
According to the Cu NMR results for Y123 at the time~\cite{Ishida}, 
Zn impurity induces a large residual Knight shift and Korringa-like behavior in the superconducting state and non-exponential nuclear spin-lattice relaxation in the normal sate, 
while Ni does not alter the normal and the superconducting properties.  
Taken together with these NMR results for Y123, the impurity effect was classified into 
\begin{eqnarray} 
\left\{
\begin{array}{l}
\mathrm{Zn: superunitary,} \\
\mathrm{Ni: subunitary.}
\end{array}
\right.
\label{BN1}
\end{eqnarray}  
The substitution of Zn impurity induces the changes in the host spin fluctuation spectrum,
while Ni does not change it but induces the depairing effect on the quasi-particle life time. 
They proposed the strong suppresssion of $T_\mathrm{c}$ 
due to a magnetic vacancy correlation function
with a vacancy operator $h({\bf r})$ of an effective radius of $\xi$:
\begin{equation}
\chi_\mathrm{Zn}({\bf r}-{\bf r'})=\chi_\mathrm{pure}({\bf r}-{\bf r'})\overline {h({\bf r})h({\bf r'})}. 
\label{SuperUnitary}
\end{equation}
The magnetic correlation is suppressed over a distance of $\xi$. 
However, this correlation function is not compatible with
all the subsequent NMR and neutron scattering results. 
 
The fact of suppression of superconductivity by nonmagnetic impurities 
has been known from early stages.  
By the Cu NMR techniques, Ishida {\it et al.} found that Zn impurity in Y123 increases the residual Knight shift 
in the superconducting state~\cite{Ishida}. 
Although they did not observe the Zn-neighbhor Cu NMR signals, 
they found that the observed Cu NMR signals are shifted by Zn impurities. 
From the residual Cu Knight shifts, they inferred a quasi-partcile energy band formation of Zn-induced virtual bound states. 
Strong scattering impurity potential in the unitarity limit must induce 
a virtual bound state in the $d$-wave superconducting state~\cite{Poilblanc,BalatskyZn,Onishi,SalkolaSTM,BulutZn,Ohashi,Ohashi1,Ohashi2}. 
Individual virtual bound states around Zn and Ni impurities are actually observed by STS technique~\cite{Pan,Hudson}. 
It remained to be solved whether the Zn
strengthens~\cite{Yamagata,AlloulBob} or weakens~\cite{Ishida} local magnetic correlation around Zn 
in the optimally doped and overdoped systems
and how the magnetic correlation near and away from Zn is changed or unchanged below and above $T_\mathrm{c}$. 
Zn-neighbor NMR and NQR measurements were highly desired. 
  
As for the optimally doped Y1237, a part of Ni impurities is substituted for the chain-site
and the amount of Ni substitution for the chain site can be controlled to some extent~\cite{AdachiNi,ItohNi}.
The Cu NQR studies of the Ni-redistributed Y1237 as well as Ni-substituted Y1248 revealed 
that Ni also induces nonexponential NMR relaxation and suppresses $T_\mathrm{c}$~\cite{ItohNi,Itoh1}. 
Further, the detailed analysis of nonexponential relaxation
revealed that no-exponential relaxation due to Zn indicates the change of the host spin fluctuation
spectrum but that due to Ni does not  necessarily indicates 
the change of the spin fluctuation spectrum~\cite{Itoh,ItohNi,Itoh1,Itoh2,Itoh3,Itoh4,Itoh5,ItohZn}.  

\subsection{SITE-SELECTIVE $^{89}$Y NMR NEAR Zn IMPURITY}
\begin{figure}[h]
\begin{center}
\includegraphics[width=0.8\linewidth]{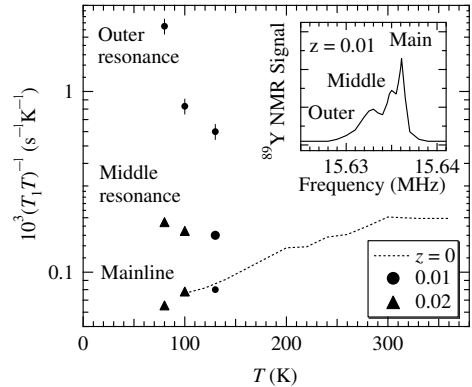}
\end{center}
\caption{
$^{89}$Y NMR spectrum (inset) and temperature dependence of
$^{89}$Y nuclear spin-lattice relaxation rate divided by temperature 1/$T_\mathrm{1}$T
for YBa$_2$(Cu$_{1-z}$Zn$_z$)$_3$O$_{6.64}$ at $T$ = 100 K and at a magnetic field applied along $ab$ plane.
All the data are reproduced from Mahajan {\it et al.}~\cite{Mahajan}.
}
\label{fig:YNMR}
\end{figure}

The local magnetic correlation is not suppressed near Zn impurity.  
Mahajan {\it et al.} succeeded in observing the $^{89}$Y NMR signals
 near Zn impurities in Zn-substituted YBa$_2$Cu$_3$O$_y$~\cite{Mahajan}.
It is an outstanding site-selective observation in the NMR studies of impurity-substituted high-$T_\mathrm{c}$ superconductors.
Figure~\ref{fig:YNMR} shows $^{89}$Y NMR spectrum (inset figure) for YBa$_2$(Cu$_{0.98}$Zn$_{0.02}$)$_3$O$_{6.6}$
and the site-selected $T_{1}$ near and away from Zn, reproduced from~\cite{Mahajan,MahajanEPJ}.   
The Zn impurity induces Curie type magnetism near the impurity site
and enhances the nuclear spin-lattice relaxation rate.  
Impurity-site NMR studies have also been intensively carried out
for Li-substituted Y123y~\cite{Mahajan,Bobroff1,MacFarlane,Bobroff2,Quazi1,Quazi2}.

\subsection{INELASTIC NEUTRON SCATTERING FOR Zn-SUBSTITUTED Y123}
Inelastic neutron scattering experiments allow us to see the dynamical spin susceptibility $\chi "({\bf q}, \omega)$
as functions of wave vector $\bf q$ and frequency $\omega$. 
Kakurai {\it et al.} observed a large change of
the dynamical spin susceptibility of Zn-substituted YBa$_{2}$Cu$_{3}$O$_{6.35}$~\cite{Kakurai}. 
Not only low energy but also high energy spectrum is changed by Zn substituion. 
Subsequent neutron scattering studies revealed for slightly-underdoped and optimally doped samples 
that not a simple shift of the spectrum weight but the development of an  in-gap state
is induced by Zn substitution~\cite{Sidis}.  

 \begin{figure}[h]
\begin{center}
\includegraphics[width=0.6\linewidth]{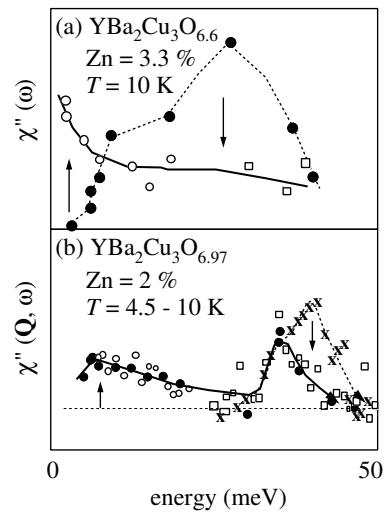}
\end{center}
\caption{
Zn-induced change in dynamical spin susceptibility  $\chi "({\bf q}, \omega)$ 
around an antiferromagnetic wave vector $\bf Q$
measured by inelastic neutron scattering experiments. 
All the data are reproduced from Kakurai {\it et al.}(a) for underdoped Y123~\cite{Kakurai} 
and Sidis {\it et al.}(b) for optimally doped Y1237~\cite{Sidis}. 
}
\label{IND}
\end{figure} 
The large change at low frequency was confirmed by a complete wipeout effect of Cu NQR signals by Kobayashi {\it et al.} for underdoped Y123~\cite{Kobayashi}. 
A partial wipeout effect by Zn impurities has already been observed by Yamagata {\it et al.} for optimally doped Y1237~\cite{Yamagata}. One should note that the relation between the in-gap state and the wipeout effect on Cu NQR 
for Zn-substitution is parallel to that for superconductor-insulator boundary.  
 
\section{ZERO FIELD Cu NQR STUDIES OF IMPURITY-INDUCED EFFECTS}
\subsection{IMPURITY-INDUCED NUCLEAR SPIN-LATTICE RELAXATION}  
In contrast to $^{89}$Y NMR, Zn-induced effect on Cu NMR and NQR spectra are
not apparently separable near and away from Zn impurity~\cite{Ishida,Walstedt,JulienZn}. 
Then, there had been a pessimistic view that no one can extract separable information
of local magnetism from Cu NMR and NQR experiments.
However, it turned out that the application of  an impurity-induced nuclear spin-lattice relaxation theory~\cite{McHenry,McHenry1}
enables us to see such local information near the impurities 
from Cu NQR~\cite{Itoh,ItohNi,Itoh1,Itoh2,Itoh3,Itoh4,Itoh5,ItohZn}.
Also, the finding of Zn-induced Cu NQR signals promotes the Cu NQR studies of 
the local antiferromagnetic correlation near the impurities~\cite{WilliamsCu,ItohZn}. 

Let us briefly explain the impurity-induced nuclear spin-lattice relaxation theory~\cite{McHenry,McHenry1}.
This theory has been applied for dilute Heisenberg insulators and dilute alloys. 
Assuming only the random distribution of impurities, a stretched exponential recovery curve
is derived for any magnetic insulators and itinerant systems, 
because it is concerned with the kinetics of nuclear moments.  
 
\begin{figure}[h]
\begin{center}
\includegraphics[width=0.6\linewidth]{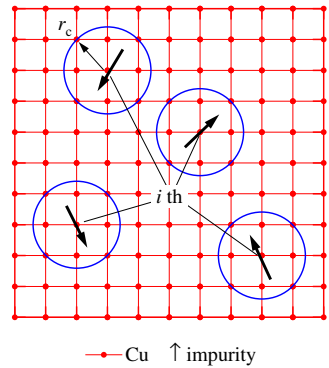}
\end{center}
\caption{
A square lattice CuO$_2$ plane with magnetic impurities.  
Solid red circles are Cu nuclear sites. 
The nuclei within a radius (wipeout radius) of  $r_{c}$ of the impurity are unobservable 
in conventional resonance frequency and time scale. 
Blue large circles indicate the wipeout regions.
}
\label{WOR}
\end{figure}

Figure~\ref{WOR} is a top view of a CuO$_2$ plane with magnetic impurities.  
Solid red circles are the Cu nuclear sites. 
The relaxation process of a nuclear moment at a $i$-site consists of two parts:
\begin{equation}
\nonumber
\mathrm{(fluctuations)_{host} + \sum(fluctuations)_{imp}},
\label{impSF}
\end{equation}
where the summation is taken over a distribution of impurities in a configuration.  
If one can observe the recovery curve $p$(${\bf r}_i$, $t$) of one nuclear spin at a $i$th site in Fig.~\ref{WOR},
it must be a single exponential function

\begin{eqnarray}
p({\bf r}_{i}, t)&=& 
\mathrm{exp}[-t\{({1\over T_{1}})_\mathrm{host}+\sum_{j}{1\over T_{1}({\bf r}_{ij})}\}] , 
%
%
\nonumber \\ &=&
\mathrm{exp}[-({t\over T_{1}})_\mathrm{host}]\prod_{j}\mathrm{exp}[-{t\over T_{1}({\bf r}_{ij})}],
\label{IINSL1}
\end{eqnarray}
where 1/$T_1(r)=A(r)^2S_\mathrm{imp}(r, \omega_n)$ is a nuclear spin-lattice relaxation rate
occurring by a nuclear-impurity coupling. 
Here, $r$ is a distance between a nuclear and an impurity-induced moment, 
$A(r)$ is a coupling constant between the nuclear spin and the impurity-induced moment, 
$S_\mathrm{imp}(r, \omega_n)$ is the impurity-induced spin-spin correlation function~\cite{Kilian,Ohashi,Ohashi1}. 
$\omega_n/2\pi$ is a nuclear resonance frequency.
For an isolated local moment on the impurity site, 
we have a longitudinal direct dipole coupling $A(r)^2\propto 1/r^6$,  
and a two-dimensional RKKY interaction $A(r)^2\propto 1/r^4$~\cite{McHenry,McHenry1}.
For an impurity-induced moment with a local $A(r)\approx A(0)$,  
we have $S_\mathrm{imp}(r, \omega_n)\propto r^{-d}$ and e$^{-r/\xi}$ 
($\xi$ is a correlation length, and $\alpha \propto d^{-1}$ is a constant)~\cite{Itoh1}.  
Instead of including all the direct and indirect nuclear-electron interactions,   
we assume a {\it minimal} model with a single interaction and 1/$T_1(r)\propto1/r^{2d}$.

The observed nuclear magnetization consists of 
nuclear spins with various configuration of randomly distributed impurities. 
The ensemble average is taken over all possible impurity configuration. 
Using a probability distribution, we obtain the nuclear magnetization   

\begin{eqnarray}
\langle p({\bf r}_{i}, t)\rangle_{\mathrm{AV}}&=&
\mathrm{exp}[-({t\over T_{1}})_\mathrm{host}]\langle \prod_{j}\mathrm{exp}[-{t\over T_{1}({\bf r}_{ij})}]\rangle_{\mathrm{AV}},
\nonumber \\ &=&
\mathrm{exp}[-({t\over T_{1}})_\mathrm{host}]
\nonumber \\ && 
\times \prod_{j}\{(1-c)+c\mathrm{exp}[-{t\over T_{1}({\bf r}_{ij})}]\},
\nonumber \\ &\approx&
\mathrm{exp}[-({t\over T_{1}})_\mathrm{host}
\nonumber \\ && 
-{c\over V}\int_{r_c}^{\infty}d^{\mathrm{D}}{\bf r}(1-\mathrm{exp}[{-t\over T_{1}({\bf r})}])],
\label{eq:IINSL2}
\end{eqnarray}
where $c$ is the impurity concentration and the observed nuclear spins are assumed to lie on a space $D$ dimension.
The nuclei within a radius (wipeout radius) of  $r_{c}$ of the impurity are unobservable 
in conventional resonance frequency and time scale, 
so that they are excluded from the integration. 
In Fig.~\ref{WOR}, blue large circles indicate the wipeout regions of a radius $r_c$.

Taken into account a finite wipeout radius $r_c$, 
the recovery curve of eq.(\ref{eq:IINSL2}) is expressed by
\begin{eqnarray}
p(t)&=&
p(0)\mathrm{exp}[-({t\over T_{1}})_\mathrm{host}
\nonumber \\ && 
- N_{c}\{\mathrm{e}^{-t/t_{c}}-1+\sqrt{\pi t\over t_{c}}\mathrm{erf}\sqrt{t\over t_{c}}\}],
\label{IINSL3}
\end{eqnarray}
where $N_c$ is wipeout number and $t_{c}$ is $T_1(r=r_c)$. 
After integrating eq.(\ref{eq:IINSL2}) 
in the limit of $r_{c}\rightarrow$0, the recovery curve is 
\begin{eqnarray}
p(t)&=&
p(0)\mathrm{exp}[-({t\over T_{1}})_\mathrm{host}-({t\over \tau_{1}})^{n}],
\nonumber \\ &=& 
p(0)\mathrm{exp}[-({t\over T_{1}})_\mathrm{host}-\sqrt{t\over \tau_{1}}],
\label{IINSL4}
\end{eqnarray}
where an impurity-induced nuclear spin-lattice relaxation rate 
\begin{eqnarray}
{1\over \tau_{1}}&\propto&
({c\over V})^{1/n}{(\gamma_{e}\gamma_{n}\hbar)^{2}S(S+1)\over \Gamma_{m}}, 
\label{tau1}
\end{eqnarray}
where $n$(=$D$/$d$) is 1/2 and $\Gamma_{m}$ is the decay rate of an impurity spin autocorrelation function. 
Thus, we obtain a theoretical recovery curve expressed by a product function of exponential function times stretched exponential function, eq.(\ref{IINSL3}) and eq.(\ref{IINSL4}).  

\begin{figure}[h]
\begin{center}
\includegraphics[width=1.1\linewidth]{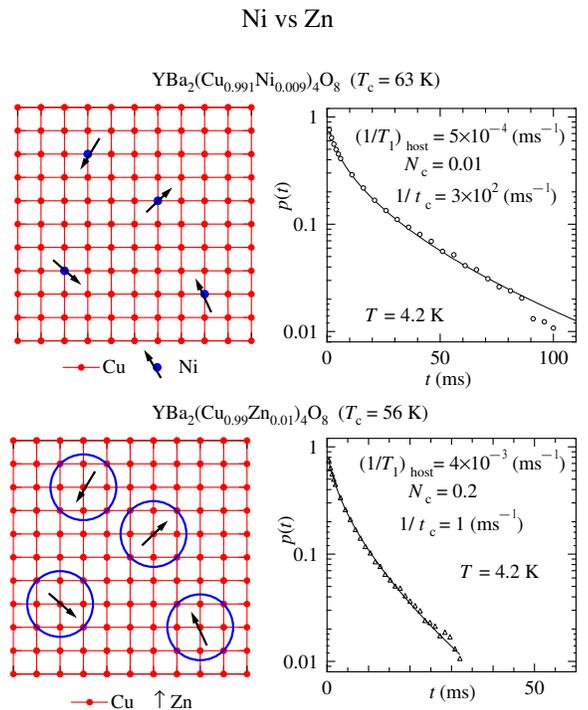}
\end{center}
\caption{
Right figures show nonexponential $^{63}$Cu(2) nuclear spin-lattice relaxation curves of Ni (upper panel) and Zn (lower panel) substituted Y1248 at $T$ = 4.2 K. 
The solid curves are least squares fits by eq.~(\ref{IINSL3}) 
based on impurity-induced NMR relaxation theory with wipeout effect.
Left figures illustrate the CuO$_2$ planes with the impurities and the wipeout regions. 
All the data are reproduced from~\cite{Itoh}. 
}
\label{WOT1}
\end{figure} 
In eq.(\ref{IINSL4}), 
$p(0)$, $(T_1)_\mathrm{host}$ and $\tau_1$ are the fit parameters. 
The original recovery curve expressed by $p(t)=p(0)\exp[-t/(T_1)_\mathrm{host}]\prod_i[(1-c)+c\exp(-t/T_1({\bf r}_i))]$
is a function of a lot of time constants of $(T_1)_\mathrm{host}$, $T_1({\bf r}_1)$, $T_1({\bf r}_2)$, $T_1({\bf r}_3)$, $\cdots$.
For $c\ll$1, the product $\prod_i[\cdots]$ is approximated by $\approx\exp[-c\int d{\bf r}(1-\exp(-t/T_1({\bf r})))]$, 
and then the spatial integral leads to the stretched exponential function with the single time constant $\tau_1$.
Thus, only two time constants, $(T_1)_\mathrm{host}$ and $\tau_1$, are obtained. 
$(T_1)_\mathrm{host}$ is the Cu nuclear spin-lattice  relaxation time 
due to the host Cu electron spin fluctuation via a hyperfine coupling. 
$\tau_1$ is the impurity-induced nuclear spin-lattice relaxation time. 
The distribution of $T_1$ is taken into consideration through $T_1(r)$
but is not known {\it a priori}.
Thus, we must assume the above-mentioned minimal form.     

These theoretical curves are reproduced by more sophisticated manner~\cite{SA}.
In passing, one may find a mathematically same expression for diffusion-limitted relaxation~\cite{Blumberg,Lowe}. 
It should be noted that the diffusion-limitted relaxation is based on the nuclear spin-spin relaxation in a long time $t$
and the impurity-induced non-exponenital function in a short time $t$. 
Thus, the application of the form of eq.(\ref{eq:IINSL2}) itself is justified for any cases.  

Figure~\ref{WOT1} shows the nonexponential $^{63}$Cu(2) nuclear spin-lattice relaxation curves 
of Ni- and Zn-substituted Y1248 at $T$ = 4.2 K and
the least squares fits using eq.~(\ref{IINSL3}) (solid curves)~\cite{Itoh}. 
In eq.(\ref{IINSL3}), 
$p(0)$, $(T_1)_\mathrm{host}$, $t_1$, and $N_c$ are the fit parameters. 
Both Ni and Zn impurities induce nonexponential recovery curves.  
A difference in Ni and Zn impurity effects is the wipeout number $N_c$ and the size of the wipeout radius $r_{c}$. 
The wipeout radius around Zn is one order larger than that around Ni~\cite{Itoh,Itoh2}. 
 
\subsection{SHRINK OF WIPEOUT REGION AROUN Zn}
\begin{figure}[h]
\begin{center}
\includegraphics[width=0.8\linewidth]{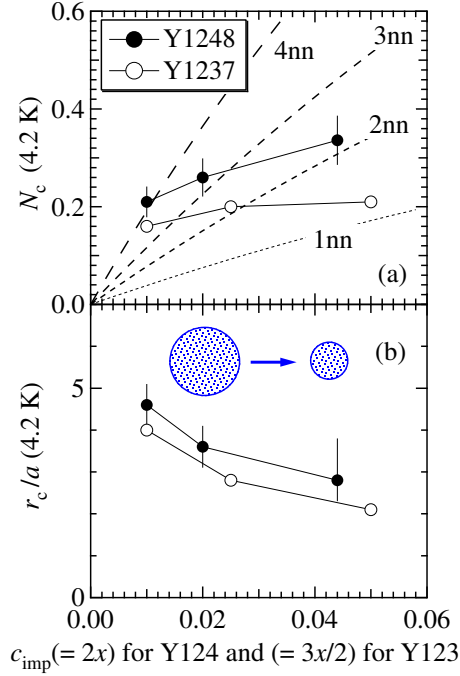}
\end{center}
\caption{
(a)Wipeout number $N_c$ and (b)wipeout radius $r_c$ as functions of Zn concentration 
for Y1237 and Y1248 at $T$ = 4.2 K~\cite{Itoh4}. 
Wipeout region per Zn shrinks with Zn substitution at $T$ = 4.2 K.
}
\label{Shrink}
\end{figure}

Figure~\ref{Shrink} shows the Zu substitution dependence
of the wipeout number $N_c$ and the estimated wipeout radius $r_c$ 
for Y1237 and Y1248 at $T$ = 4.2 K~\cite{Itoh4}. 
The wipeout region around Zn shrinks as Zn concentration is increased. 
This is associated with the Zn-induced change of host antiferromagnetic spin fluctuations. 

The size of the local magnetic state induced around Zn decreases as Zn concentration is increased, 
because the host antiferromagnetic correlation length $\xi_\mathrm{AF}$ decreases with Zn (see below). 
On the other hand, 
the superconducting virtual bound state (Andreev bound state) is induced around Zn
and its size increases as Zn concentration is increased, 
because the upper critical field $H_\mathrm{c2}$ decreases with Zn and 
the superconducting coherence length $\xi_\mathrm{SC}$ increases with Zn~\cite{Tomimoto}. 
A virtual bound state around Zn is actually observed by STS technique~\cite{Pan}. 
The core radius is estimated to be $\approx\xi_\mathrm{SC}$. 

In the theory based on SO(5) symmetry, 
if $\xi_\mathrm{SC}$ is equal to $\xi_\mathrm{AF}$, 
the coexistence of superconductivity and antiferromagnetic long range ordering could be realized~\cite{Zhang,Kohno}. 
In an underdoped regime, Zn recovers the coherency of magnetic correlation 
over the destructive effect of magnetic dilution~\cite{Hucker}. 
Then, the effect of Zn on the size of the magnetic bound state may not be monotonic. 
Such recovery of the N{\' e}el ordering was also observed 
for La$_{2-x}$Sr$_x$CuO$_4$ with Ni~\cite{Machi}.    

\subsection{Zn VS Ni}
\begin{figure}[h]
\begin{center}
\includegraphics[width=1.1\linewidth]{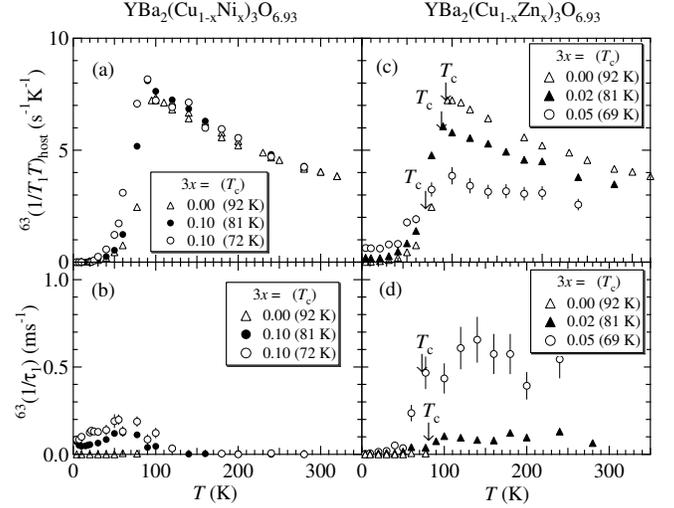}
\end{center}
\caption{
Host $^{63}$Cu(2) nuclear spin-lattice relaxation rate (1/$T_{1}T$)$_\mathrm{host}$ and 
impurity-induced $^{63}$Cu(2) nuclear spin-lattice relaxation rate 1/$\tau_{1}$
for optimally doped Y1237 with Ni impurity (a) and (b)~\cite{ItohNi} and with Zn impurity (c) and (d)~\cite{ItohZn,Itoh5} .
}
\label{Y123W}
\end{figure}
\begin{figure}[h]
\begin{center}
\includegraphics[width=1.1\linewidth]{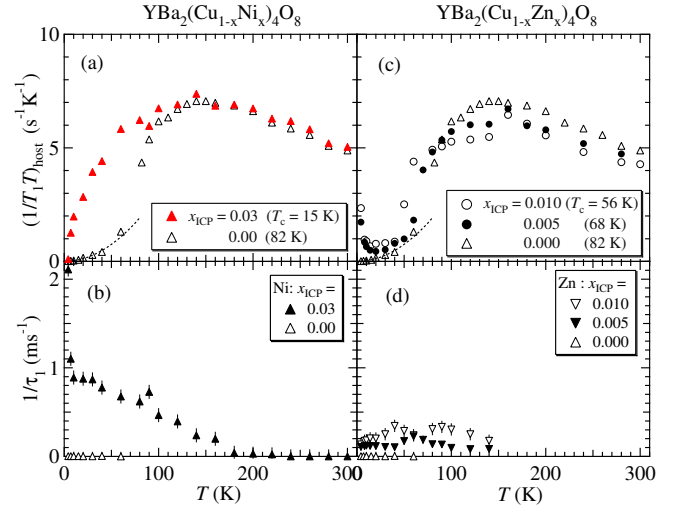}
\end{center}
\caption{
Host $^{63}$Cu(2) nuclear spin-lattice relaxation rate (1/$T_{1}T$)$_\mathrm{host}$ and 
impurity-induced $^{63}$Cu(2) nuclear spin-lattice relaxation rate 1/$\tau_{1}$ 
for naturally underdoped Y1248 with Ni impurity (a) and (b)~\cite{Itoh1,Itoh} and with Zn impurity (c) and (d)~\cite{Itoh3,ItohZn}. 
}
\label{Y124W}
\end{figure}
Although we believe that eq.(\ref{IINSL3}) with a finite $N_c$ reproduces the actual recovery curve
for impurity-substituted samples better than eq.(\ref{IINSL4}) with a fixed $N_c$ = 0~\cite{Itoh2}, 
there is a shortcoming of overestimation of $N_c$ above $T_\mathrm{c}$~\cite{Itoh}. 
Thus, we adopt the next best policy and apply eq.(\ref{IINSL4}) to obtain host and guest spin dynamics. 

Figure~\ref{Y123W} shows (1/$T_{1}T$)$_\mathrm{host}$ and 1/$\tau_{1}$ using eq.~(\ref{IINSL4}) 
for Ni- and Zn-substituted optimally doped Y1237~\cite{ItohNi,ItohZn,Itoh5}.
Figure~\ref{Y124W} shows (1/$T_{1}T$)$_\mathrm{host}$ and 1/$\tau_{1}$ using eq.~(\ref{IINSL4}) 
for Ni- and Zn-substituted underdoped Y1248~\cite{Itoh1,Itoh,Itoh3,ItohZn}. 
Both Ni and Zn induce nonexponential nuclear spin-lattice relaxation for the underdoped and the optimally doped samples.
But, for low impurity concentration, the host spin fluctuation is robust  for Ni substitution but fragile for Zn substitution.
The pseudo spin-gap behavior of 1/$T_1$ is robust both for Ni and Zn substitution. 
For high impurity concentration, the strong wipeout effects on Cu NQR signals and 
the suppression of pseudo spin-gap are observed for both Ni- and Zn-substituted underdoped Y1248~\cite{Itoh3}.
This is consistent with the neutron scattering measurement for Zn-substituted oxygen deficient Y123 
by Kakurai {\it et al}~\cite{Kakurai}.

\subsection{Cu(2) NQR NEAR Zn}  
Zn impurities induce low frequency satellite signals in the plane-site Cu(2) NQR spectra
for Zn-substituted Y1248~\cite{WilliamsCu,Itoh3,ItohZn} and Y1237~\cite{Yamagata,ItohZn}.  
From the relative intensity of Cu NQR spectra, 
the satellite signals are assigned to Zn-neighbor Cu NQR lines.
The Cu nuclear spin-lattice relaxation time of the satellite signal is shorter than 
that of the main signal in the superconducting state of Y1237 and Y1248
and even above $T_\mathrm{c}$ for Y1248~\cite{ItohZn}.  
Although Y1248 is a stoichiometric, homogenous, underdoped electronic system,  
the Zn-induced inhomogeneous magnetic response in the CuO$_2$ plane   
is more marked than that of the optimally doped Y1237.  

\begin{figure}[h]
\begin{center}
\includegraphics[width=1.1\linewidth]{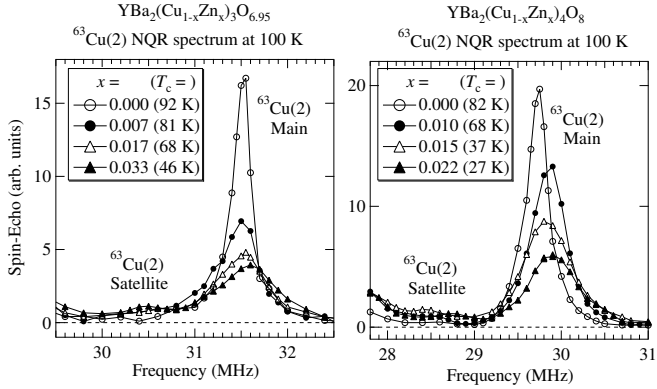}
\end{center}
\caption{
Zn-induce satellite signals and reduced main signals of $^{63}$Cu NQR spectra for Y1237 and Y1248 
reproduced from Itoh {\it et al.}~\cite{ItohZn}.
}
\label{ZnNQR}
\end{figure}  
\begin{figure}[h]
\begin{center}
\includegraphics[width=1.1\linewidth]{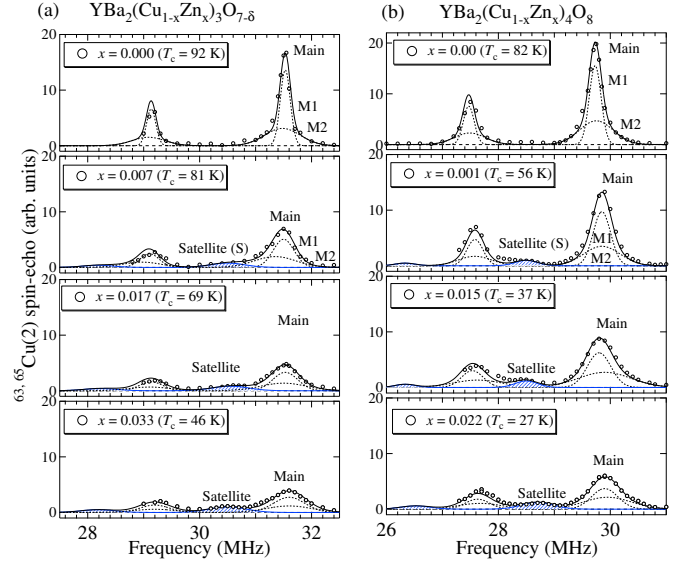}
\end{center}
\caption{
Zn-substitution effects on the plane-site Cu NQR spectra for Y1237 and Y1248 at 100 K~\cite{ItohZn}.
Individual NQR spectra are fit by three Gaussian functions (dashed curves) of M1, M2, and S. 
Original integrated Cu NQR intensity of each pure sample is given by $I$(0) = M1 + M2
for Zn-free Y1237 and Y1248. 
}
\label{ZnCuNQR}
\end{figure} 
\begin{figure}[h]
\begin{center}
\includegraphics[width=1.1\linewidth]{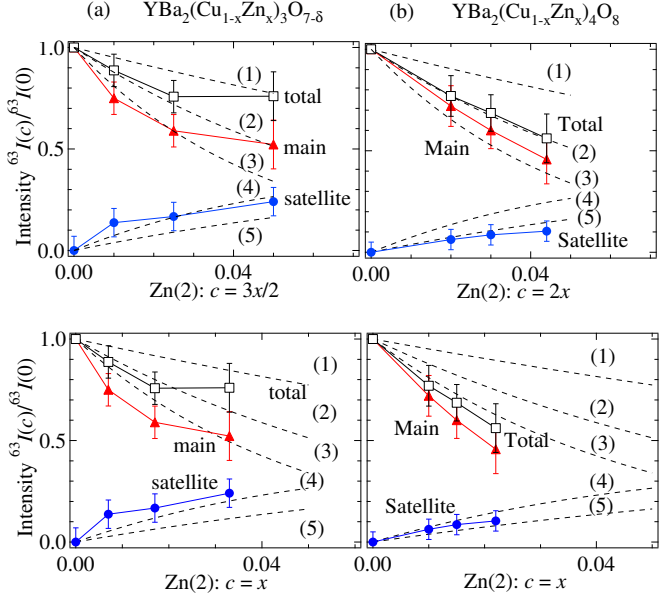}
\end{center}
\caption{
Zn-substitution dependence of integrated Cu NQR spectra using multiple Gaussian functions for Y1237 and Y1248 at 100 K.
Red triangles are the main signal intensity. 
Blue circles are the satellite signal intensity. 
Open squares are the summation of the main and satellite signals. 
The dashed curves are numerical calculations using binominal distribution functions.  
The in-plane Zn concentration $c$ is assumed to be $c$ = 3$x$/2 for Y1237 and $c$ = 2$x$ for Y1248 in upper panels,
while in lower panels, $c$ = $x$ is assumed both for Y1237 and Y1248.
}
\label{IntCuNQR}
\end{figure} 
Figure~\ref{ZnNQR} shows the Zn impurity substitution effects on $^{63}$Cu(2) NQR spectra 
at $T$ = 100 K in the normal states of Y1237 (a) and Y1248 (b).
Figure~\ref{ZnCuNQR} shows the $^{63,65}$Cu(2) NQR spectra at $T$ = 100 K  
for Y1237 (a) and for Y1248 (b) from the top to the bottom as functions of Zn concentration.  
The main signal intensity of the $^{63}$Cu(2) NQR spectrum around 31.5 MHz for Y1237 and around 29.8 MHz for Y1248 
rapidly decreases as Zn impurity concentration is increased. 
The lower frequency tail and the broad satellite signal increases as Zn impurity concentration is increased.
The solid and the dotted curves are numerical simulations using three Gaussian functions ($M_1$, $M_2$, and $S$). 
By two Gaussian functions of $M_1$ and $M_2$, we can approximate the main spectra. 
The shaded Gaussian $S$ is the satellite spectrum. 

In Fig.~\ref{IntCuNQR}, 
the relative intensity of the main spectrum $I_\mathrm{main}$(=$M_1$+$M_2$) (upward solid triangles), 
the satellite spectrum $I_\mathrm{sate}$(=$S$) (open circles),
and the total intensity(=$M1$+$M2$+$S$)
are plotted as functions of the plane-site Zn(2) concentrations 
for Y1237 (a) and Y1248 (b). 
All the $I_\mathrm{main}$, $I_\mathrm{sate}$ and the total intensity are normalized by $I(0)$.  
In the upper panels, all Zn impurities are assumed to be substituted for the plane-site Cu(2);
$c$ = 3$x$/2 for Y1237 and $c$ = 2$x$ for Y1248.  
In the lower panels, the Zn impurities are assumed to be equally substituted both for the chain-site Cu(1) and plane-site Cu(2);
$c$ = $x$ for Y1237 and Y1248. 

The dashed curves are numerical calculations of Cu NQR intensity with various Zn configurations, 
using the binomial distribution function  
\begin{eqnarray}
_{n(j)}B_k(c)\equiv_{n(j)}C_kc^k(1-c)^{n(j)-k},
\label{IINSL1}
\end{eqnarray}
where $n(j)$ is the number of the $j$th nearest neighbor (nn) Cu sites.  
(1-$c$) is the probability that a plane site is occupied by a Cu ion but not a Zn ion, 
that is, the total Cu NQR intensity with Zn substitution.
With respect to Zn configuration, the probability of finding Cu atoms 
can be seen in the decomposition of  
\begin{eqnarray}
1 &=&
1_\mathrm{1st}\cdot 1_\mathrm{2nd}\cdot 1_\mathrm{3rd}\cdots 
\nonumber \\ &=&
\prod_{j=1}1_{j\mathrm{th}}
\nonumber \\ &=&
\prod_{j=1}\{(1-c)+c\}^{n(j)}
\nonumber \\ &=&
\prod_{j=1}\sum_{k=0}^{n(j)}{_{n(j)}B_k(c)}.
\label{binomials}
\end{eqnarray}

The total number of probability of finding Cu decreases as Zn concentration $c$ is increased,
that is (1-$c$). Keeping the lowest order terms of $c$($\ll$ 1),  we obtain
\begin{eqnarray}
(1-c) &=&
(1-c)\prod_{j=1}\sum_{k=0}^{n(j)}{_{n(j)}B_k(c)}
\nonumber \\ &=&
(1-c)[(1-c)^{\sum_{j=1}^N n(j)}
\nonumber \\ &&
+\sum_{j=1}^N n(j)c(1-c)^{n(j)-1}+O(c^2)].
\label{binomials2}
\end{eqnarray}

\begin{figure}[h]
\begin{center}
\includegraphics[width=1.0\linewidth]{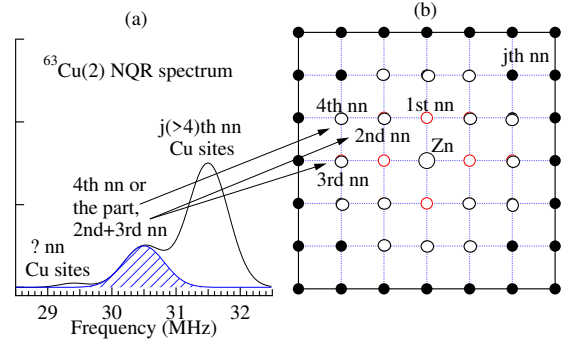}
\end{center}
\caption{
Illustrations of (a) a $^{63}$Cu(2) NQR spectrum and (b) a top view of the CuO$_2$ plane with a Zn impurity
for Zn-substituted Y1237.
The shaded area in (a) is assigned to the Zn-neighbor Cu NQR signals.
In (b), Zn locates at the center, open circles are the $j$th nn Cu sites near Zn,
and solid circles are Cu sites away from Zn.  
}
\label{CuO2Zn}
\end{figure}
Figure~\ref{CuO2Zn} illustrates a $^{63}$Cu(2) NQR spectrum (a) and 
the CuO$_2$ plane with a Zn impurity (b) for Zn-substituted Y1237.
The shaded area in (a) is  the Zn-induced Cu NQR signal.
In (b), Zn locates at the center, open circles are the $j$th nn Cu sites near Zn,
and solid circles are Cu sites away from Zn.  
These are guides for the below calculations of NQR signal intensity. 

The calculated curves on the Cu NQR intensity are in what follows.  
In Figs.~\ref{IntCuNQR}, the decreasing functions as increasing $c$ are 
\begin{eqnarray}
I(c)/I(0)&=&
\left\{
\begin{array}{l}
(1-c)_4B_0(c)=(1-c)^5 (1),  \\
(1-c)(_4B_0(c))^3=(1-c)^{13} (2), \\
(1-c)(_4B_0(c))^4=(1-c)^{21} (3).
\end{array}
\right.
\label{BN1}
\end{eqnarray}  
These are the probability of finding the Cu atoms with all Cu atoms in their 1st nn positions (1), 
in their 12 neighboring (up-to 3rd nn) positions (2), and in their 20 neighboring (up-to 4th nn) positions (3). 
In other words, the diminished Cu NQR signals come from the Cu atoms with at least one Zn atom 
in their 1st nn positions (the 1st nn wipeout effect) (1), 
in their up-to 3rd nn positions (wipeout effect up to the 3rd nn sites) (2), and 
in their up-to 4th nn positions (wipeout effect up to the 4th nn sites) (3). 

In Fig.~\ref{IntCuNQR}, the increasing functions as increasing $c$ are 
\begin{eqnarray}
I(c)/I(0)&=&
\left\{
\begin{array}{l}
(1-c)_8B_1(c)=8c(1-c)^8 (4), \\
(1-c)_4B_1(c)=4c(1-c)^4 (5). 
\end{array}
\right.
\label{BN2}
\end{eqnarray} 
These are the probability of finding the Cu atoms with one Zn atom in their 4th nn positions {\it or} in the
2nd and 3rd nn positions (4), and in their $j$(=1, 2, 3)th nn positions (5).

In the upper panel of Fig.~\ref{IntCuNQR}, 
the Zn-doping dependence of the main intensity for Y1237 ($x\leq$0.017) and Y1248 ($x\leq$0.022) 
is close to that of the case (3) within the experimental accuracy, 
where the Cu sites located up to the 3rd nn from Zn are unobservable (wipeout). 
The Zn doping dependence of the total intensity for Y1237 ($x\leq$0.017) and Y1248 ($x\leq$0.022) 
is close to that of the case (2), 
where the Cu sites located up to the 4th nn from Zn are unobservable (wipeout).  
For Y1237, the satellite signal is close to either of the Cu site at the 4th nn from Zn  
{\it or} the Cu sites at the 2nd and 3rd nn from Zn [the case (4)]. 
For $x\leq$0.017, the former assignment is consistent with those to the total and the main signals, 
whereas for $x$=0.033, the latter is consistent with those to the total and the main ones.     
For Y1248, the satellite signal is close to the Cu site at the 1st, 2nd, or 3rd nn from Zn [the case (5)].
However, this is inconsistent with the wipeout effect on the total intensity up to the 3rd nn. 
Thus, the satellite resonance of Y1248 may arise from the fraction of the Cu site at the 4th nn from Zn. 

The different assignment of the satellite signal from the above~\cite{ItohZn} is reported in~\cite{WilliamsCu}, 
where the satellite is assigned to the 1st nn Cu site near Zn. 
The point that the satellite comes from Zn-neighbhor some Cu sites is consistent with each other. 
The wipeout effect on the Cu NQR main signals and the growth of the Zn-induced Cu satellite signals 
indicate that the satellite but not the main signal involves the Zn-neighbor Cu nuclei. 
However, another site assignment is proposed from theoretical calculation using cluster model,
where the higher frequency side of the main Cu NQR signal involves the 1st nn Cu sites
and the satellite signal is assigned to the 2nd nn Cu site to Zn~\cite{Meier}.
No wipeout effect is supposed. 
The actual Cu NQR spectrum does not consist of sharp split lines but rather broad bands.
That is a reason which makes it difficult to assign the signals. 

\subsection{Cu(2) NUCLEAR SPIN-LATTICE RELAXATION TIME NEAR Zn}
The substitution of Zn impurity for Cu site broadens the Cu NQR spectrum
and induces the low frequency satellite signals. 
Nuclear spin-lattice relaxation results from spin fluctuations through local nuclear sites in a real space~\cite{Moriya0}.
We are concerned about frequency distribution of Cu nuclear spin-lattice relaxation time
and the temperature dependence. 
The detailed measurements of frequency dependence of Cu nuclear spin-lattice relaxation curves
revealed that the spin fluctuations depend on the distance from Zn impurity. 
 
\begin{figure*}[t]
\begin{center}
\includegraphics[width=0.7\linewidth]{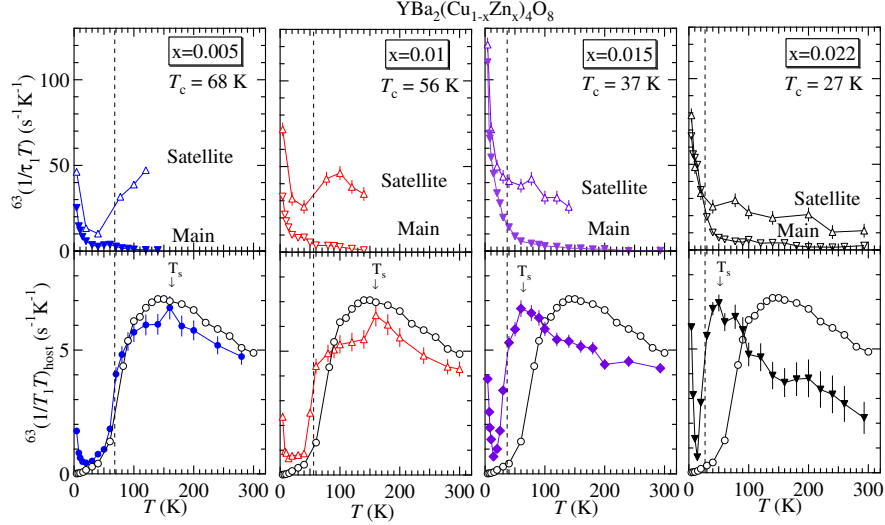}
\end{center}
\caption{\label{T1ZnY124}
Zn-substitution effects on $^{63}$Cu(2) nuclear spin-lattice relaxation rates
at Cu NQR main and satellite signals
$^{63}(1/\tau_1T$) and $^{63}(1/T_1T)_\mathrm{host}$ 
as functions of temperature for for Y1248.  
Open circles are $^{63}(1/T_1T)_\mathrm{host}$ for pure Y1248. 
The dashed lines indicate the respective $T_\mathrm{c}$'s.
$T_s$ is the pseudo spin-gap temperature defined by the maximum temperature of $^{63}(1/T_1T)_\mathrm{host}$.
All the data are reproduced from~\cite{ItohZn,Itoh3}
 }
\end{figure*}

Figure~\ref{T1ZnY124} shows $^{63}(1/\tau_1T)$ and $^{63}(1/T_1T)_\mathrm{host}$ 
of the satellite and main signals 
as functions of temperature for Zn-substituted Y1248, respectively~\cite{ItohZn,Itoh3}. 
The relaxation rates were estimated by using eq.(\ref{IINSL4}).
The main signals have both components of $^{63}(1/\tau_1T)$ and $^{63}(1/T_1T)_\mathrm{host}$,
whereas the satellite signals have a single  component of $^{63}(1/\tau_1T)$.
For Y1237, the satellite signals have both components~\cite{ItohZn}.  
The difference in $^{63}(1/\tau_1T)$ between Y1237 and Y1248 
is associated with the degree of underlying magnetic correlation and
the size of the pseudo spin-gap. 

In the superconducting state, $^{63}(1/\tau_1T)$ increases as temperature is decreased.
The difference in the temperature dependence of $^{63}(1/\tau_1T)$ between the main and the satellite signals
is marked even above $T_\mathrm{c}$.  
The main $^{63}(1/T_1T)_\mathrm{host}$ above about 100 K decreases with Zn substitution.
For low Zn concentration, the pseudo spin-gap temperature $T_s$ is invariant,
while for high Zn concentration, $T_s$ rapidly decreases~\cite{ItohZn,Itoh3}. 
The collapse of pseudo spin-gap  
suggests the occurrence of a quantum phase transition.  
In the underdoped regime, Zn impurity induces superconductor-insulator crossover or transition.

The Cu(2) nuclear spin-lattice relaxation time probes 
the in-plane antiferromagnetic dynamical spin susceptibility at an NMR/NQR frequency
~\cite{Shastry,MMP,MTU,Bulut}
and reflects homogeneity of the CuO$_2$ plane.
Relation between the Cu NQR relaxation results~\cite{ItohZn,Itoh3} and the $^{89}$Y NMR results~\cite{Mahajan}  
is not straightforward.  
 
$^{63}(1/\tau_1T)$ of the satellite signal is enhanced more than that of the main signal. 
This is an evidence for the Zn-induced virtual bound state~\cite{Balatsky,Onishi,Salkola} 
via the locally enhanced magnetic correlation~\cite{Ohashi,Ohashi1,BulutZn,Poilblanc}.    
The difference in $^{63}(1/\tau_1T)$ of the satellite from the main indicates 
that Zn-induced ``staggered moments" persist both above and below $T_\mathrm{c}$~\cite{Ohashi,Ohashi1,Ohashi2,BulutZn},  
where spatially extended potential scattering by Zn plays a key role~\cite{Balatsky1,Xiang},
and that the pseudo spin-gap effect is suppressed near Zn~\cite{YanaseZn}.    

\section{NMR STUDIES OF VORTEX CORES}
\subsection{REDFIELD PATTERN}
Not only the impurity-induced magnetism but also magnetism inside vortex cores in the mixed states 
of high-$T_\mathrm{c}$ superconductors have attracted great interests~\cite{Zhang}. 
Localized~\cite{Caroli} and delocalized quasi-particle sates~\cite{Volovik} are induced
inside the vortex cores in conventional and unconventional superconductors, respectively. 
Vortex core magnetism and vortex core charge have been explored from microscopic viewpoints~\cite{Kishine,Ogata,HIM}.  
By the STS techniques, scanning conductance spectra were measured around vortex cores 
and revealed the anomalous local density of states around the cores, indicating the existence of internal electric and magnetic structures~\cite{Fischer,Nishida}.    

\begin{figure}[h]
\begin{center}
\includegraphics[width=1.0\linewidth]{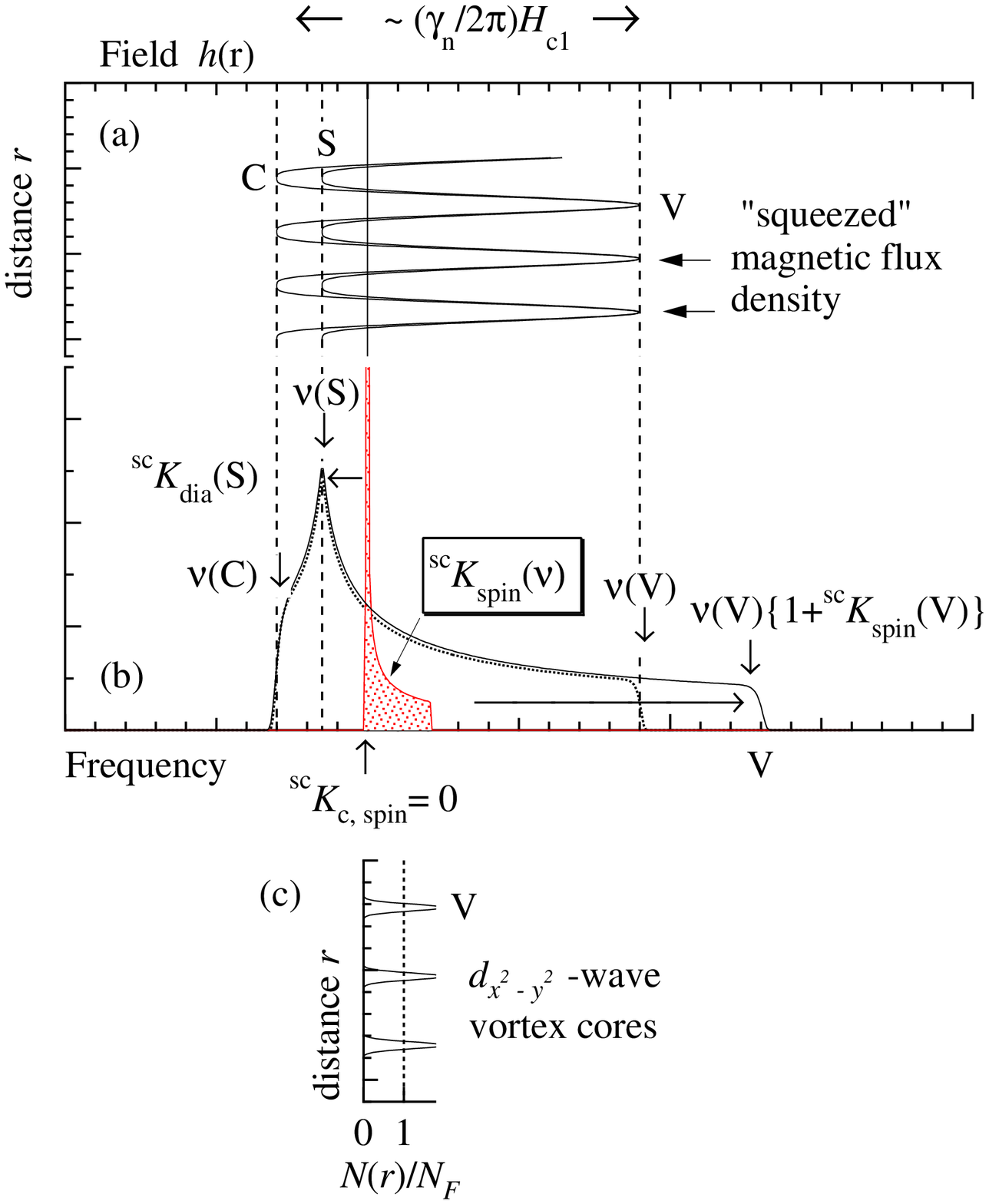}
\end{center}
\caption{
Illustrations of magnetic field distribution (a), Redfield pattern of NMR spectrum (b), 
and distribution of Knight shift due to $d_{x^2-y^2}$- wave vortex cores (c)~\cite{ItohTL}. 
Red shaded area indicates ``Redfield pattern" of distributed Knight shift.  
One should note that magnetic field distribution due to supercurrent is at most $\sim H_\mathrm{c1}$. 
}
\label{Redfield}
\end{figure} 
An NMR spectrum is known to show a characteristic pattern
in the mixed states of type-II superconductors, that is Redfield pattern~\cite{Red}. 
This results from magnetic field distribution of quantized magnetic flux produced 
by local supercurrent against an external magnetic field.  
Quantized magnetic flux forms a triangular lattice or a square lattice.  
Each quantized magnetic flux has a vortex core.
The magnetic flux spreads over a distance of a penetration depth $\lambda_\mathrm{SC}$, 
while the vortex core spreads over a distance of 
the superconducting coherence length $\xi_\mathrm{SC}(\ll \lambda_\mathrm{SC}$)~\cite{Sonier}.

At the early stage, there had been a pessimistic view that 
NMR technique cannot detect the Redfield pattern for high-$T_\mathrm{c}$ superconductors, 
because of  the long penetration depth $\lambda_\mathrm{SC}$.
However, the problem is not whether $\lambda_\mathrm{SC}$ is long or short. 
In general, the width of Redfield pattern is at most the lower critical field of $H_\mathrm{c1}$
in a dilute vortex lattice~\cite{Brandt,Brandt1,Brandt2,Rigamonti}. 
Since $\mu$on spin rotation measurements indicate $H_{c1}\propto T_\mathrm{c}$~\cite{Sonier},
the Redfield pattern can be expected for higher $T_\mathrm{c}$ compounds. 
The problem is whether the linewidth $\nu_\mathrm{NMR}$ of NMR spectrum in a normal state 
is sharp enough to detect additional broadening in the vortex state~\cite{ZPNQR}. 
We have a criterion of
\begin{equation}
\Delta \nu_\mathrm{NMR} \leq {\gamma_\mathrm{n}\over 2\pi}H_\mathrm{c1},
\label{Hc1}
\end{equation}
where $\gamma_\mathrm{n}$ is nuclear gyromagnetic ratio. 

Figure~\ref{Redfield} illustrates internal magnetic field distribution $h(r)$ due to quantized magnetic flux (a), 
Redfield pattern of NMR spectrum (b), 
and distribution of Knight shift $K_\mathrm{spin}(\nu)$ 
due to $d$-wave local density of states (c)~\cite{ItohTL}. 

An external magnetic field
is {\it squeezed} at the vortex core~\cite{Brandt,Brandt1,Brandt2}. 
The {\it squeezed} magnetic flux density at the core
produces a higher magnetic field than an applied magnetic field,
so that the shift at the nuclear site inside the core is larger than the shift in a normal state.

Local density of states of electrons recovers at the vortex cores 
of $d$-wave superconductors~\cite{TIM}. 
Then, not only the distribution of an applied magnetic field 
but also the distribution of Knight shift due to anisotropic local density of states 
contributes the NNR pattern~\cite{ItohTL}.  

Using a vortex core shift 
$^{sc}K_\mathrm{c, field}$(V) $> $0 due to the enhanced field and
a negative shift 
$^{sc}K_\mathrm{c, dia}$(C)$ < $0 due to the diamagnetic field,
we obtain the resonance frequency $\nu(\mathrm{V})$ inside and $\nu(\mathrm{C})$ outside the vortex cores, 

\begin{equation} 
\nu(\mathrm{V})=
\nu_{0}[1+^{sc}K_\mathrm{c, spin}(\mathrm{V})][1+^{sc}K_\mathrm{c, field}(\mathrm{V})],
\label{e.core}
\end{equation}
and
\begin{equation} 
\nu(\mathrm{C})=\nu_{0}[1+^{sc}K_\mathrm{c, dia}(\mathrm{C})].
\label{e.mini}
\end{equation}
One should take into consideration
the combined effect of magnetic flux squeezing and of the finite positive spin shift at the vortex core~\cite{ItohTL}. 

The Redfield pattern was actually observed in type-II superconductors, e.g.
vanadium~\cite{Red}, LiTi$_2$O$_4$~\cite{MItoh}, and MgB$_2$~\cite{MgB2A,MgB2B}.  
In the high-$T_\mathrm{c}$ cuprate superconductors, 
the Redfield patterns were also observed, e.g.
by $^{17}$O NMR of optimally doped Y1237 and underdoped Y1248~\cite{Curro,Reyes,Bachman,Mitro,Kaku}
and $^{209}$Tl NMR of overdoped Tl$_2$Ba$_2$CuO$_{6+\delta}$ (Tl2201) ($T_\mathrm{c}$ = 85 K)~\cite{Mehring,Kumagai,ItohTL}. 
The frequency distribution of $^{17}T_1$ in the Redfield pattern was observed 
for the first time by Curro {\it et al} for Y1237~\cite{Curro}. 
This is the beginning of site-selective NMR studies of vortex core magnetism. 

In passing, the evidence of static antiferromagnetic vortex cores at $T_\mathrm{N}$ = 20 K and at $H$ = 2 T
was reported for Tl2201~\cite{Kumagai}. 
But, the irreversible magnetization line was identified at $T_\mathrm{irr}$ = 20 K and at $H$ = 2 T,
and the slowing down effect of vortex melting or solidification was observed 
in $^{209}$Tl NMR~\cite{ItohTL}.  
From magnetic field dependence of $^{209}T_1$, 
a direct process of overdamped motion of pancake vortices was reported
to play a key role~\cite{pancake}. 
Whether the static antiferromagnetic ordering takes place inside vortex cores,
one needs further studies.

As to the plane-site Cu(2) NMR, the ideal condition of eq.(\ref{Hc1}) is not satisfied for Y1237~\cite{TakigawaVL}. 
However, in the attempts to detect the spatial distribution of antiferromagnetic spin correlation,
we obtain a few results of Cu NMR studies of the vortex cores.  

\subsection{ZEEMAN-PERTURBED Cu NQR}
Let us present Zeeman-perturbed Cu NQR studies of vortex cores 
for optimally doped Y1237 at a low magnetic field of $H$ = 96 mT~\cite{ZPNQR}. 
The plane-site Cu(2) NQR spectrum is known to be one of the sharpest 
among the reported NQR spectra of other cuprate superconductors.  
The Cu(2) nuclei are coupled with nearly uniaxial electric field gradients~\cite{ShimizuEQQ,PenningtonEQQ}.
At high magnetic fields, various vortex pinning effects are observed 
from measurements of hysteresis curves of magnetization~\cite{Erba}. 
At low magnetic fields near $H_{c1}$, however, 
a large diamagnetic response is observed in any case. 
The vortex lattice at a low magnetic field is in a quasi-long range ordering state~\cite{Koba}.  
In the Cu NMR experiment for Y1237, 
the low magnetic field yields only a negligibly small Cu Knight shift~\cite{TakigawaKs} 
but is expected to cause a large diamagnetic shift. 

\begin{figure}[h]
\begin{center}
\includegraphics[width=0.8\linewidth]{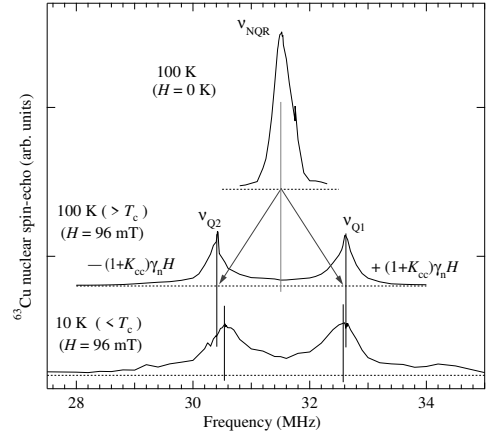}
\end{center}
\caption{
From top to bottom:
zero field $^{63}$Cu (nuclear spin $I$ = 3/2) NQR spectrum at $T$ = 100 K,
Zeeman-split $^{63}$Cu NQR spectra at $T$ = 100 K and then 10 K 
at an external magnetic field of $H$ = 96 mT along the $c$-axis~\cite{ZPNQR}.  
The Zeeman-split directions of $^{63}$Cu NQR lines (solid lines with $\nu_{\mathrm{Q1}}$ and $\nu_{\mathrm{Q2}}$) 
are denoted by arrows.   
The diamagnetic shift is superimposed on the original broad NQR lines.  
}
\label{ZPNQR}
\end{figure}

Figure~\ref{ZPNQR} shows a zero-field $^{63}$Cu NQR power spectrum at 100 K ($> T_\mathrm{c}$) and 
the Zeeman-split NQR spectra at 100 K and 10 K ($< T_\mathrm{c}$).  
The magnetic field of $H$ = 96 mT was applied along the $c$ axis, 
the maximum principle axis of electric field gradients. 
An intervortex distance at 96 mT is about 1800 $\AA$, 
longer than the penetration depth 
$\lambda_{ab}$($T\rightarrow$0 K) = 1100-1300 $ \AA$~\cite{Sonier}. 
$H_{c1}$ is estimated to be about 110 mT at $T\rightarrow$0 K~\cite{Hc1}.

Magnetic field splits the $^{63}$Cu NQR spectrum with a peak frequency $\nu_{\mathrm{NQR}}$
into two lines with lower and higher frequency peaks 
($\nu_{Q1}$ and $\nu_{Q2}$)~\cite{DasHahn,TakigawaLFNMR}.
By definition of $^{63}\gamma_n^{\ast}$$\equiv$(1+$K_\mathrm{cc}$)$^{63}\gamma_n$ 
($^{63}\gamma_n$ = 11.285 MHz/T and the $c$-axis Knight shift $K_\mathrm{cc}$),
the transition frequencies of $\nu_{Q1}$ ($I_z$=3/2$\leftrightarrow$1/2) and 
$\nu_{Q2}$ ($I_z$=$-$3/2$\leftrightarrow$$-$1/2) are given by 
 
\begin{eqnarray}
\left\{
\begin{array}{l}
\nu_{\mathrm{Q1}}=\nu_{\mathrm{NQR}}+^{63}\gamma_{n}^{\ast}H,\\
\nu_{\mathrm{Q2}}=\nu_{\mathrm{NQR}}-^{63}\gamma_{n}^{\ast}H.
\end{array}
\right. 
\label{eq.ZPNQR}
\end{eqnarray}
One should note that $K_\mathrm{cc}$ involves both the Knight shift and the superconducting diamagnetic shift. 
From $\nu_{Q1}$ and $\nu_{Q2}$,  
one can estimate the quadrupole frequency $\nu_\mathrm{NQR}$($H$) at a magnetic field $H$ 
and the local magnetic field $\delta h$[$\equiv$(1+$K_\mathrm{cc}$)$H$] 
by

\begin{eqnarray}
\left\{
\begin{array}{l}
\nu_{\mathrm{NQR}}(H)=(\nu_{Q1}+\nu_{Q2})/2, \\
\delta h=(\nu_{Q1}-\nu_{Q2})/2{}^{63}\gamma_{n}.
\end{array}
\right. 
\label{eq.eqQKs}
\end{eqnarray} 
The Cu NQR spectrum at 100 K is broadened by the distribution of electric field gradients.  
The split $^{63}$Cu NQR lines is broadened at 10 K larger than 100 K. 
The local field due to the Knight shift~\cite{TakigawaKs} 
is negligible at the magnetic field of $H$ = 96 mT.  
Thus, the additional broadening at 10 K must be due to the superconducting diamagnetic shift
$^{\mathrm{sc}}K_{c, \mathrm{dia}}$ and 
the distribution of the electric field gradient $\nu_{\mathrm{NQR}}(H)$ in a mixed state.  

\begin{figure}[h]
\begin{center}
\includegraphics[width=0.8\linewidth]{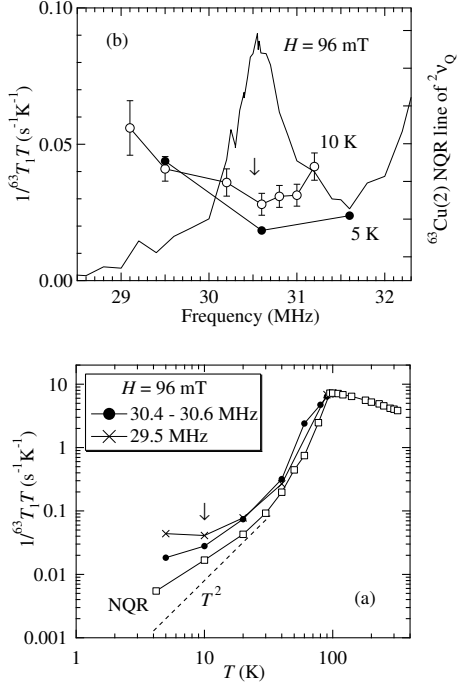}
\end{center}
\caption{
(a) Frequency distribution of
$^{63}$Cu nuclear spin-lattice rate divided by temperature 1/$^{63}T_{1}T$ at $T$ = 10 and 5 K
(left axis) in the Zeeman-perturbed $^{63}$Cu(2) NQR spectrum of $\nu_{Q2}$ (right axis)~\cite{ZPNQR}. 
The arrow indicates a minimum of 1/$^{63}T_{1}T$
for nuclei close to the saddle points in a vortex lattice. 
(b) Temperature dependence of 
the $^{63}$Cu nuclear spin-lattice relaxation rate divided by temperature 1/$^{63}T_{1}T$ 
across the split NQR spectrum~\cite{ZPNQR}.
The arrow indicates a shallow minimum of 1/$^{63}T_{1}T$ at 10 K
for the 29.5 MHz nuclei close to the vortex cores.
}
\label{ZPNQRT1}
\end{figure}
From the estimated 31.54 $\leq\nu_{\mathrm{NQR}}(\mathrm{96 mT})\leq$31.58 MHz 
and $\nu_{\mathrm{NQR}}$(0) = 31.54 MHz at 10 K,
we obtain the upper limit of the field-induced quadrupole frequency 
\begin{equation} 
0\leq[\nu_{\mathrm{NQR}}(96 \mathrm{mT})-\nu_\mathrm{NQR}(0)]\leq 40 \mathrm{kHz}. 
\label{eq.DeqQ}
\end{equation}
The low magnetic field of 96 mT does not enhance the vortex charge,
compared with that of 9.4 T~\cite{VCKuma}.
There is no robust evidence of vortex charging effect on NQR frequency 
beyond the experimental accuracy. 

Figure~\ref{ZPNQRT1}(a) shows the frequency dependences of 
the $^{63}$Cu nuclear spin-lattice relaxation rate divided by temperature 
1/$^{63}T_{1}T$ at $T$ = 10 and 5 K in the Zeeman-perturbed $^{63}$Cu(2) NQR spectrum of $\nu_{Q2}$ (right axis). 
Above $T_\mathrm{c}$, 1/$^{63}T_{1}$ shows no appreciable frequency dependence,
whereas below $T_\mathrm{c}$, it shows a strong dependence. 
This is an evidence for the site-selective measurement of $^{63}T_{1}$ 
in the Zeeman-split NQR spectrum and for the less effect of nuclear spin diffuison.

The frequency dependence of 1/$^{63}T_{1}$ 
is similar to those of the plane-site $^{17}$O(2, 3) and the apical $^{17}$O(4) 
at high magnetic fields of $H$ = 9-37 T~\cite{Curro,Mitro}. 
The plane-site Cu directly probes the in-plane antiferromagnetic correlation.
The frequency dependence of 1/$^{17}T_1$ is primarily explained  
by Doppler shift in quasiparticle excitation spectrum around the vortex cores~\cite{TIM,Morr}. 
The supercurrent around a vortex core induces Doppler shift 
in the local quasiparticle energy spectrum~\cite{deGennes}. 
Thus, the frequency dependence of 1/$^{63}T_{1}$, 
similar to that of the high field 1/$^{17}T_{1}$,
indicates the importance of the Doppler shift in quasiparticle energy spectrum and
the absence of the static antiferromagnetic vortex cores at 96 mT.  

Figure~\ref{ZPNQRT1}(b) shows 1/$^{63}T_{1}T$ 
as a function of temperature across the split NQR spectrum of $\nu_{Q2}$ and at zero field NQR. 
The signals at 30.4-30.6 MHz and at 29.5 MHz in the split NQR spectrum come from the nuclei
close to the saddle points and close to the vortex cores, respectively. 
The dashed line indicates 1/$T_{1}T\propto T^{2}$ of $d$-wave pairing. 
The temperature dependence of 1/$^{63}T_{1}T$ at each site 
can be understood primarily by the spatial dependence of 
the local density of states with the Dopple shift~\cite{TIM}.   
The minimum of 1/$^{63}T_{1}T$ at 10 K for nuclei at 29.5 MHz
is not understood simply by the local density of states.
  
Two theoretical explanations have been proposed.
One is the competition mechanism in the scattering of quasi-particle and antiferromagnetic spin fluctuations~\cite{Morr}.
The other is the mismatch effect of angular dependent local density of states around a vortex core
and the geometry of magnetic flux lattice~\cite{Knapp}
     
In the presence of a supercurrent,
the local antiferromagnetic spin fluctuations have two fold contributions to 1/$^{63}T_{1}T$: 
one is a scattering of a Bogoliubov particle at a spin-fermion vertex,
and the other is simultaneous creation or annihilation of two Bogoliubov particles~\cite{Morr}. 
Competition of two processes yields minimum behavior 1/$^{63}T_{1}T$ 
as functions of temperature and frequency. 

The periodicity of vortex lattice, i.e. triangular or square lattice, depends on the magnitude of an external magnetic field $H$. 
Mismatching between the magnetic flux shape and the local density of states of the quasi-particles
yields the characteristic dependence of temperature and frequency~\cite{Knapp}. 

\section{CHEMICAL PRESSURE, PHYSICAL PRESSURE, AND SITE DISORDER}
It has been believed that the site disorder out of CuO$_2$ planes
does not cause serious damage to the electronic states of the CuO$_2$ planes. 
However, the reason why chemical pressure
effect on $T_\mathrm{c}$ and the electronic states by element substitution is not the same 
as the physical pressure one has been often ascribed to the disorder effect. 
The relation among site disorder, chemical pressure and physical pressure effects 
has been poorly underdtood. 
 
Y(Ba$_{1-x}$S$_{x}$)$_{2}$Cu$_4$O$_8$ with $x \leq$ 0.40 can be synthesized~\cite{Wada1}. 
For Y1248, one can expect two effects of Sr substitution for Ba sites 
without change of oxygen content; 
one is chemical pressure and the other is crystalline potential disorder. 
The size of Sr$^{2+}$ ion is smaller than that of
Ba$^{2+}$ by about 10 $\%$, so that the substituted Sr ions make local strains 
and decrease the volume. 
The lattice constants actually shrink with Sr substitution~\cite{Wada1},
similarly to physical pressure effect~\cite{Yamada}.
Thus, the Sr substitution introduces crystalline potential
disorder and chemical pressure to the lattice. 

The effect of hydrostatic physical pressure on the spin dynamics is similar to 
the carrier doping effect~\cite{Machi1,Machi2,Machi3}: 
The physical pressure effects on $T_\mathrm{c}$ and on the pseudo spin-gap temperature $T_s$ 
agree with the carrier doping effects for Y1248, 
$T_\mathrm{c}$ increases~\cite{Kaldis,Scholtz} but $T_s$ decreases~\cite{Machi3}, 
as well as with doping Ca~\cite{Miyatake1,Machi1}.

For Y1248, the Sr substitution for Ba site scarcely increases $T_\mathrm{c}$~\cite{Wada1}, 
although the unit cell volume shrinks. 
Thus, one can suspect randomness effect on the actual
$T_\mathrm{c}$ as well as nonmagnetic impurity Zn substitution effect. 
\begin{figure}[h]
\begin{center}
\includegraphics[width=1.1\linewidth]{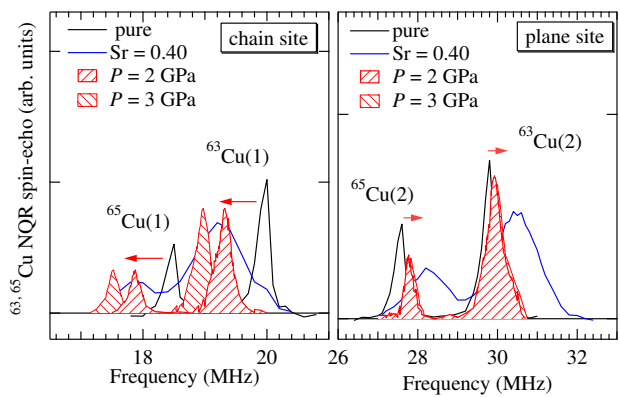}
\end{center}
\caption{
Zero field $^{63, 65}$Cu NQR frequency spectra of the chain-site Cu(1) (left side) and  the plane-site Cu(2) (right side)
for Y(Ba$_{1-x}$Sr$_x$)$_2$Cu$_4$O$_8$ of $x$ = 0.00 and 0.40 
at an ambient pressure of $P$ = 0.1 MPa~\cite{ItohYSr}. 
The red shaded Cu(1, 2) NQR spectra were measured 
under hydrostatic physical pressure of $P$ = 2 and 3 GPa~\cite{Machi2}.
}
\label{YSr}
\end{figure}

Figure~\ref{YSr} shows $^{63, 65}$Cu NQR spectra of 
the plane-site Cu(2) (a) and the chain-site Cu(1) (b) for $x$ = 0.00 and 0.40 at $T$ = 4.2 K~\cite{ItohYSr}. 
For comparison, the red shaded $^{63, 65}$Cu NQR spectra for a pure Y1248 
under hydrostatic physical pressure of $P$ = 2.0 and 3.0 GPa are also presented~\cite{Machi2}. 
By Sr substitution for Ba site, the Cu(2) NQR spectra
are shifted to higher frequencies, whereas the Cu(1) NQR spectra are shifted to lower
frequencies. 
Since the directions of these shifts are in parallel to those under the
physical pressure~\cite{Zimmermann1,Machi2}. 
However, one should note that the degree of shifts of the Cu(2) NQR spectra is different from that of
Cu(1) between the Sr-substitution and the physical pressure effects~\cite{Zimmermann1,Machi2}.
This difference indicates that local compression due to the internal pressure of Sr is different
from that due to the physical pressure.  

Both the line widths of Cu(1) and Cu(2) spectra are broadened by Sr substitution
and then the disorder of the crystalline potential due to Sr in BaO layer.
 For such a broadened spectrum, one would expect an inhomogeneous local density of electron states and an inhomogeneous electron spin dynamics. 
 However, no impurity effect on nuclear spin-lattice relaxation was observed~\cite{ItohYSr}. 
 Both Cu(1) and Cu(2) nuclear spin-lattice relaxation curves were single exponential functions. 
 Also, no Curie term in uniform spin susceptibility was observed~\cite{ItohYSr}. 
 Theses results are sharply in contrast to the effects of in-plane impurity substitution.  
It remains to be a mystery why the chemical pressure of Sr substitution does not increase $T_\mathrm{c}$. 

\section{CONCLUSION}
\label{sec:conclusion}
	Local electronic states near the in-plane impurities and inside the vortex cores have been studied 
by Cu NQR and NMR techniques for high-$T_\mathrm{c}$ superconductors Y123y and Y1248. 
The characteristic NQR and NMR spectra and the frequency distribution of Cu nuclear spin-lattice relaxation time 
provide us with rich information of magnetic correlation near Zn impurity and inside the vortex cores.       

\acknowledgments
I would like to thank Y. Ohashi, M. Ogata and Y. Yanase for their stimulating discussions of theoretical treatments, 
and T. Machi, N. Watanabe, S. Adachi, C. Michioka, and K. Yoshimura 
for their collaboration for a long time.   	 


\end{document}